\documentclass[pdflatex,sn-nature]{sn-jnl}

\usepackage{graphicx}%
\usepackage{multirow}%
\usepackage{amsmath,amssymb,amsfonts}%
\usepackage{amsthm}%
\usepackage{mathrsfs}%
\usepackage[title]{appendix}%
\usepackage{xcolor}%
\usepackage{textcomp}%
\usepackage{manyfoot}%
\usepackage{booktabs}%
\usepackage{algorithm}%
\usepackage{algorithmicx}%
\usepackage{algpseudocode}%
\usepackage{listings}%
\usepackage{siunitx}

\theoremstyle{thmstyleone}%
\theoremstyle{thmstyletwo}%

\theoremstyle{thmstylethree}%

\begin{document}

\title[Article Title]{Light Coils: MRI with Fully Optical Data and Power Transmission}

\author*[1]{\fnm{Zining} \sur{Liu}}\email{zining.liu@uniklinik-freiburg.de}
\equalcont{These authors contributed equally to this work.}

\author*[2]{\fnm{Morteza} \sur{Teymoori}}\email{morteza.teymoori@imtek.uni-freiburg.de}
\equalcont{These authors contributed equally to this work.}

\author[1]{\fnm{Jakob} \sur{Gerlach}}\email{jakob.gerlach@uniklinik-freiburg.de}

\author[1]{\fnm{Reza} \sur{Aghabagheri}}\email{reza.aghabagheri@uniklinik-freiburg.de}

\author[3]{\fnm{Henning} \sur{Helmers}}\email{henning.helmers@ise.fraunhofer.de}

\author[1]{\fnm{Michael} \sur{Bock}}\email{michael.bock@uniklinik-freiburg.de}

\author[2]{\fnm{Çağlar} \sur{Ataman}}\email{caglar.ataman@imtek.uni-freiburg.de}

\author[1]{\fnm{Ali Caglar} \sur{Özen}}\email{ali.oezen@uniklinik-freiburg.de}

\affil[1]{\orgdiv{Department of Diagnostic and Interventional Radiology}, \orgname{University of Freiburg}, \orgaddress{\street{Killianstr. 5a}, \city{Freiburg}, \postcode{79110}, \state{Baden-Württemberg}, \country{Germany}}}

\affil[2]{\orgdiv{Department of Microsystems Engineering}, \orgname{University of Freiburg}, \orgaddress{\street{Georges Köhler-Allee 102}, \city{Freiburg}, \postcode{79110}, \state{Baden-Württemberg}, \country{Germany}}}

\affil[3]{\orgdiv{Department III-V Photovoltaics and Concentrator Technology}, \orgname{Fraunhofer Institute for Solar Energy Systems ISE}, \orgaddress{\street{Heidenhofstr. 2}, \city{Freiburg}, \postcode{79110}, \state{Baden-Württemberg}, \country{Germany}}}

\abstract{In magnetic resonance imaging (MRI) dense receiver coil arrays with a high number of coil elements are used to efficiently detect and encode the signal. Further increasing the number of coils is hampered by electrical cabling and massive electronics that introduce electromagnetic coupling, integration complexity and even safety constraints. Here we introduce the novel Light Coils concept, a fully optical MRI receive architecture, in which data transmission, front-end power delivery and coil detuning are all implemented optically, which allows reducing the massive galvanic cabling to a few optical fibers. For signal encoding, Mach–Zehnder modulators (MZM) are used to convert the preamplified RF signal from each coil onto a C-band optical carrier. The preamplifiers are driven via a power-over-fiber (PoF) system that uses a high-efficiency photovoltaic (PV) cell for optical-to-electrical power conversion. A pulse-sequence-triggered optical path controls active detuning. Jointly optimizing modulator bias, optical power and front-end gain under realistic receiver chain conditions, Light Coils can match the signal-to-noise ratio (SNR) of conventional RF coil systems with galvanic cables at MZM input powers of 5–10\,\unit{\mW} and photonic power converter inputs of 80-100\,\unit{\mW}. At a clinical 3 Tesla MRI system we show \textit{in vivo} human brain imaging with a single-channel Light Coil element with an image quality and SNR comparable to a conventional coaxial readout using the identical coil element. Extending the concept to a four-channel array using dense wavelength-division multiplexing over a single fiber, we demonstrate wavelength-selective routing with inter-channel optical isolation exceeding 28\,\unit{\decibel}, reduced noise correlation compared with the galvanic reference, and parallel imaging. These results establish a scalable route towards lightweight, modular, and potentially ultra-dense MRI receive arrays based on integrated photonics and power-over-fiber.
}


\maketitle

\section{Introduction}\label{intro}
Magnetic resonance imaging (MRI) is a widely used non-invasive imaging modality that provides cross-sectional anatomical images with excellent soft-tissue contrast and functional information. It is also one of the most powerful tools available for investigating human brain structure and function. A persistent limitation of contemporary research and clinical MRI systems is the slow data acquisition, arising from limitations in gradient-based spatial encoding and signal-to-noise ratio (SNR). The resulting long acquisition times on the order of minutes are challenging for subjects and increase the risk of motion-induced artifacts.

Recent advances such as parallel imaging \cite{pruessmann1999sense,griswold2002generalized}, compressed sensing \cite{lustig2008compressed}, and deep learning \cite{yaman2020self, Cukur2025} have accelerated MRI by enabling reconstruction from undersampled data. SNR can be increased by using contrast agents \cite{caravan1999gadolinium}, higher-field systems \cite{ocali1998ultimate}, or indirectly using improved gradient coil hardware \cite{foo2020highly}, but improvements in receiver coil technology often provide greater SNR gains at a lower cost \cite{roemer1990nmr,wiesinger2004electrodynamics,corea2016screen, Bock2006ActiveCatheterTracking} and are directly compatible with existing MRI systems. In particular, dense receiver arrays support high acceleration factors with reduced g-factors (i.e., an SNR penalty due to noise correlation among receive elements), thereby enabling high spatial and temporal resolution. Arrays with up to 64 channels are now widely used. Notably, in a recent study, the potential contribution of a simulated 256-channel coil was estimated and shown to be superior to a commercial 32-channel coil by a factor of 1.5 in the image SNR and 2.5 in the acceleration factor \cite{hendriks2019potential}.

Although it offers major advantages in imaging performance, increasing coil density introduces fundamental engineering and safety challenges. The main limitation is the extensive cabling required to route the MR signals to the receiver and supply electrical power to the preamplifiers and active detuning circuits. These cables interact with electromagnetic fields and can induce electric fields in the patient, leading to unwanted RF-induced heating. Cable traps and BALUNs mitigate these effects, but add bulk, rigidity, and weight to the system. Densely packed cables also degrade SNR and parallel imaging performance through cross-talk between adjacent lines, with typical isolation between neighboring coaxial cables falling below 20\,\unit{\decibel}. As a result, coil arrays beyond 64 channels are difficult to design, manufacture, and deploy using conventional approaches \cite{schmitt2008128,tokuda2009lung,uugurbil2019brain,wiggins200996,hardy2008128}, and performance compromises often offset potential benefits.

Several strategies have been proposed to address these limitations. Wireless transmission schemes can transfer analog or digital MR signals \cite{aggarwal2016millimeter,wei2007realization,byron2017rf}, but require amplification, digitization, compression, and active detuning, all of which still depend on electrical power typically delivered via coaxial lines. Optical data transmission for MRI has also been demonstrated \cite{simonsen2019magnetic, tang2015home, fandrey2012novel, memis2008miniaturized, nobre2022optical, simonsen2019sensitive, gupta2012digital, fandrey2008development, yuan2007direct,yuan20084, koste2005magnetic}. While these illustrated the potential of optical communication for MRI, they relied on individual laser sources in each coil element and still used conductive wiring for power delivery, which limits scalability. 

Power-over-fiber (PoF) is a technology for delivering power over non-conducting, non-magnetic optical fibers \cite{matsuura2021}, thus being particularly useful for sensing and communication systems, where electrical isolation and electromagnetic interference (EMI) immunity are primary concerns \cite{kamiyama2018multichannel, putra2023, al-zubaidi2021}. As such, it can be adapted for MRI receive coil arrays for powering the on-coil electronics, such as low-noise amplifiers (LNA) and detuning circuits. To the best of our knowledge, this work reports the first MRI receiver coil array architecture (Light Coils) that employs PoF and demonstrates a unified approach that provides both data transmission and power delivery via optical means. 

\begin{figure}
    \centering
    \includegraphics[width=\linewidth]{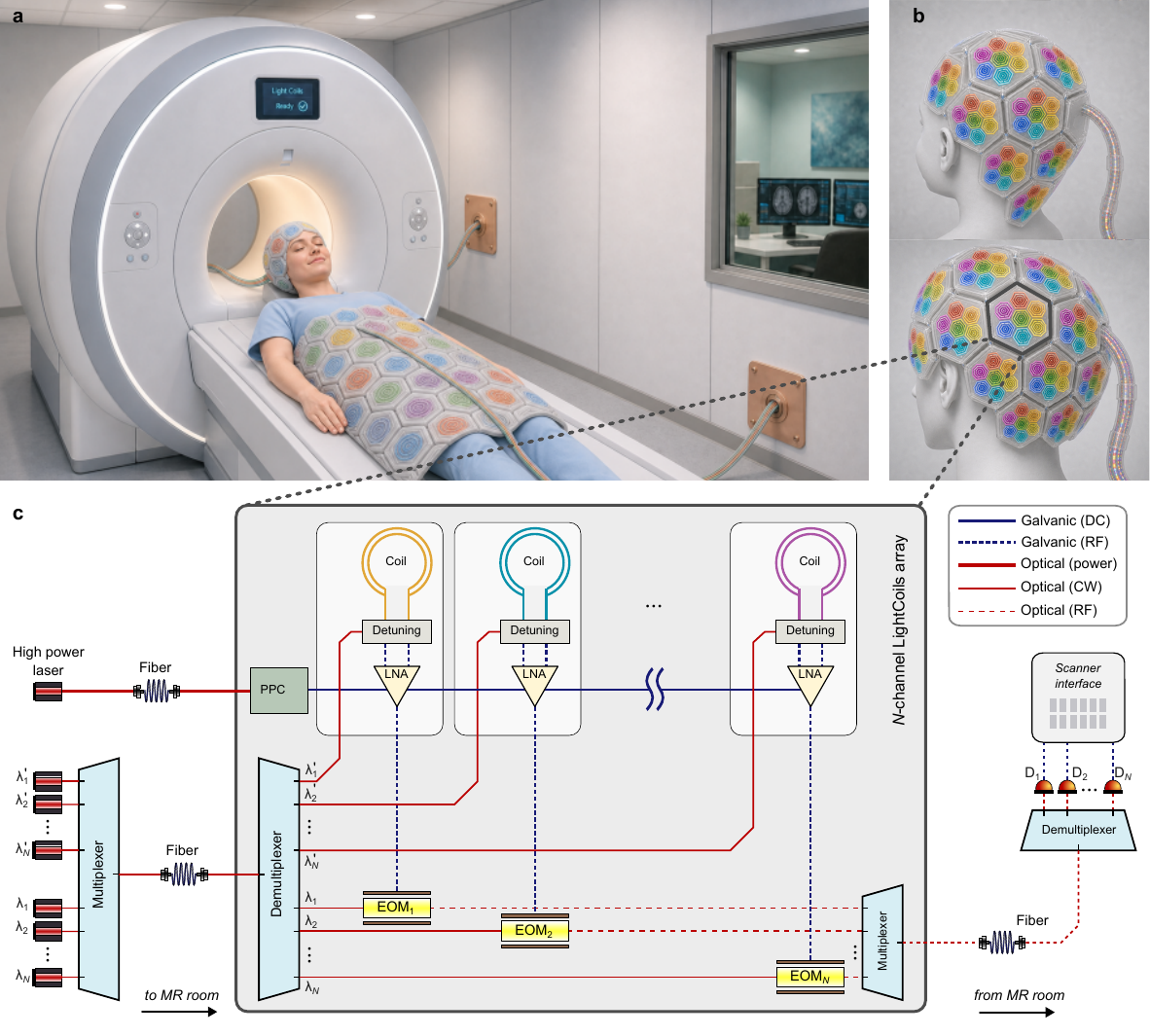}
    \caption{\textbf{Conceptual description of the Light Coils approach. (a)} Conceptual clinical implementation of Light Coils as lightweight, optically connected receive-array modules placed close to the patient inside the MRI scanner. Optical fibers route signal, power, and control connections between the patient-side coil array and the scanner-side interface, reducing conventional galvanic cabling around the patient. \textbf{(b)} Example modular array-of-arrays configurations for brain imaging. Individual Light Coil modules or sub-arrays can be tiled around the head to increase anatomical coverage and adapt the receive geometry to different imaging targets or patient sizes. \textbf{(c)} Each Light Coil module comprises two main units separately located in the control and MRI rooms, which are connected through optical fiber cables only. High-power laser is delivered to photonic power converters (PPCs) to supply the on-coil low-noise amplifiers (LNAs), while data transmission and detuning control use dense wavelength division multiplexing for individual control of the coil elements. The MR signal from each coil is amplified and converted to optical signal by electro–optic modulator (EOM). The modulated optical signals are recombined, transmitted out of the MRI room, demultiplexed and converted back to electrical signals.}
    \label{fig1}
\end{figure}

In Figure \ref{fig1}, we outline the general concept of Light Coils as a lightweight and reconfigurable “array-of-arrays” architecture, which uses PoF and dense wavelength division multiplexing (DWDM) as central design elements. As illustrated in Figure \ref{fig1}a and b, by combining RF coil technologies with intrinsic decoupling, individual coil elements and pre-assembled coil modules can be tiled to form larger, application-specific arrays, such as dense head arrays or extended body arrays, while maintaining channel independence and scalability. This modularity allows tailoring the coil array geometry and coverage to each individual patient and target anatomy; for example, different numbers of modules can be used for imaging the head, spine, torso, or even small organs such as the prostate. The underlying technical concept is shown in Figure \ref{fig1}c. Instead of using separate conductive cables for RF, power, and control-signal transmission, the Light Coil architecture transmits the MR signal, front-end power, and active detuning control solely via optical pathways. Distinct optical wavelengths can be assigned to different receive elements or sub-arrays by WDM components, so that multiple MR signals can be transmitted through the same optical fiber. Additional optical channels can provide PoF to a high-efficiency photonic power converter (PPC) for LNA powering and wavelength-selective control for active detuning \cite{helmers20206,helmers202168}. After electro-optic conversion at the coil, the optical signals are routed out of the Faraday's cage of the MRI room to the outside MRI receiver electronics. Here, they are converted back into the electrical domain to be compatible with the existing MRI system architecture, which digitizes the signals for later image reconstruction. In general, the Light Coils concept simplifies magnet-side electronics and dramatically reduces the need for conductive cabling, thus avoiding unwanted noise coupling and potentially dangerous cable heating.

In this work, we first demonstrate the key building blocks of Light Coils experimentally. We characterize an analog optical link based on a Mach–Zehnder modulator and identify the operating conditions that maximize image SNR under realistic conditions. We then integrate optical power delivery via a PPC and sequence-synchronized optical detuning, enabling a single-channel Light Coil that achieves imaging performance comparable to a conventional galvanic reference. We further extend the architecture to a four-channel data-only Light Coils array built on a DWDM. The isolation between channels is evaluated to assess scalability. Phantom and \textit{in vivo} MRI experiments demonstrate the capabilities of parallel imaging. To our knowledge, this is the first proof of concept for simultaneous optical data and optical power transmission in an MRI receive array, opening a path towards dense, modular receiver coils beyond the current state of the art.

\newpage
\section{Results}\label{result}

\subsection{Optical data transmission}
We first characterized the analog optical link independently of the optical detuning and optical powering subsystems. The analog optical link consisted of an LNA followed by a Mach–Zehnder modulator (MZM), which converts the amplified MR voltage signal onto a 1550\,\unit{\nm} optical carrier. As shown in Figure \ref{fig2}a, in the bench characterization setup, an RF signal source replaced the coil and LNA. The modulated optical signal was transmitted to a high-speed photodetector, where it was converted back into an RF signal for subsequent reception by the MRI system’s receiver interface. A detailed description of the link architecture and a small-signal model describing gain, noise and linearity as functions of MZM bias and input optical power are provided in Supplementary Note 1.

\subsubsection{Bench tests}
\begin{figure}[ht!]
    \centering
    \includegraphics[width=1\linewidth]{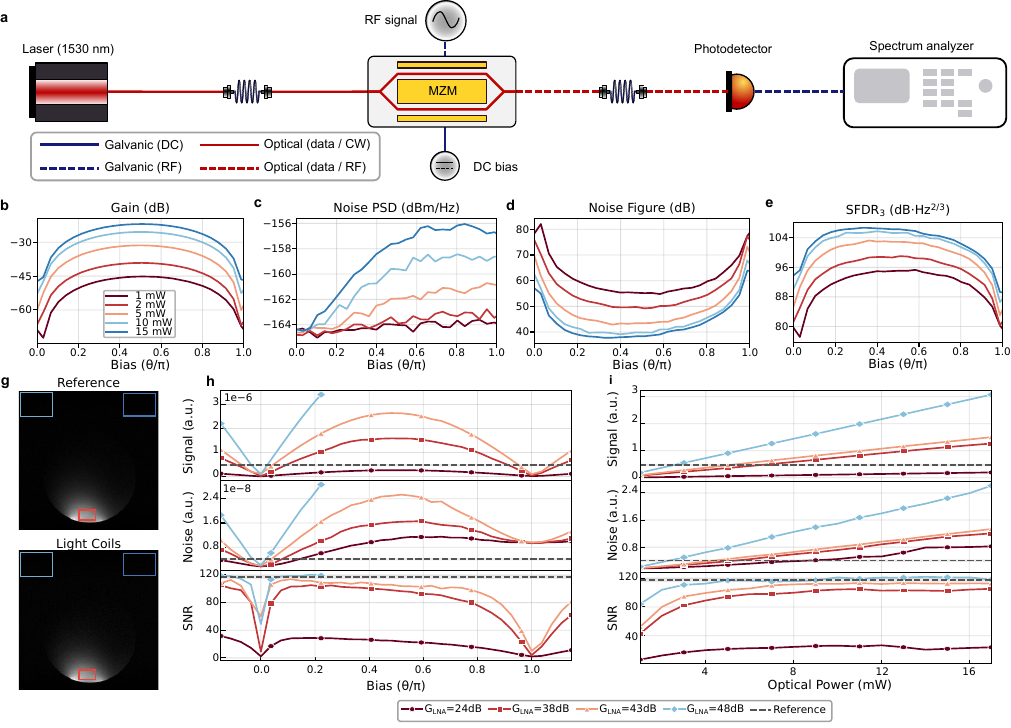}
    \caption{\textbf{Characterization of the analog optical data transmission platform for the Light Coil. (a)} Schematic of the analog optical link. (b-e) Bench characterization of the analog optical link at 123.2\,\unit{\MHz} as a function of MZM bias point ($\theta/\pi$), shown for laser powers of 1, 2, 5, 10, and 15\,\unit{\mW}: \textbf{(b)} link gain, \textbf{(c)} output-referred noise power spectral density (PSD), \textbf{(d)} noise figure (NF), and \textbf{(e)} third-order spurious-free dynamic range (SFDR\textsubscript{3}). \textbf{(g)} axial GRE phantom images acquired using galvanic transmission (top) and optical data transmission (bottom), and the corresponding signal and noise ROI selection for SNR calculation. \textbf{(h)} Bias-point sweep at a fixed laser power of 10\,\unit{\mW}, showing signal mean, noise standard deviation, and SNR (mean/STD) extracted from the ROIs in (g) for low-noise amplifier (LNA) gains of 24, 38, 43, and 48\,\unit{\decibel}. GLNA denotes the gain of the LNA placed directly after the coil and before the electro-optical conversion stage of the analog optical link. The dashed line labeled "Reference" indicates the SNR obtained with the galvanically connected reference. \textbf{(i)} Corresponding laser-power sweep at a fixed bias point of $\theta=0.2\theta$, showing signal, noise, and SNR versus optical power for the same LNA gain settings, with the galvanic reference again shown as a dashed line.}
    \label{fig2}
\end{figure}

Figure \ref{fig2}b–e summarizes the measured analog optical link performance as a function of the normalized MZM bias angle $\theta$ for five optical input powers between 1 and 15\,\unit{\mW}. As expected for an analog MZM, the link gain (Figure \ref{fig2}b) followed a $\mathrm{sin}^2(\theta)$ envelope, peaking at quadrature ($\theta$\,=\,$\pi/2$) and reaching nulls at $\theta$\,=\,$0$ and $\theta$\,=\,$\pi$, where the modulator is biased at the extrema of its transfer function. Increasing the input optical power of the MZM shifted the gain curves upward across the full bias range, consistent with the increased optical power available for RF modulation and photodetection. These measurements confirm that both the MZM bias and the optical input power effectively control the optical link gain, with quadrature bias maximizing small-signal gain.

The output noise power spectral density exhibited a more complex dependence on the MZM bias angle (Figure \ref{fig2}c). At low input optical power (1\,\unit{\mW}), the noise floor was constant at –164\,\unit{\decibel\milli\per\hertz}, dominated by the spectrum analyzer's noise floor. As the optical power increased, the noise floor rose near the quadrature point and remained low close to the transfer function nulls (i.e. $\theta$\,=\,$0,\,2\pi$), reflecting an increasing contribution from relative-intensity noise (RIN) of the laser and shot noise of the photodetector, both of which scale with the optical power. The bias dependence of gain and noise can be understood by considering that the link gain follows a $\mathrm{sin}^2(\theta)$ envelope, whereas shot noise and RIN scale linearly and quadratically, respectively, with the average optical power, which itself has a $1-\mathrm{cos}(\theta)$ dependency. Consequently, a modest reduction in link gain can be accompanied by a larger reduction in link-added noise, resulting in a bias point with a lower noise figure (NF) away from the gain maximum (Figure \ref{fig2}d). The calculated third-order spurious-free dynamic range (SFDR\textsubscript{3}) is shown in Figure \ref{fig2}e; it followed the bias dependence of the NF and remained high over the low-noise operating region.  Increasing the input optical power improved SFDR\textsubscript{3}, with the highest measured value reaching approximately 105\,\unit{\decibel \raiseto{2/3} \hertz} at 15\,\unit{\mW}. This indicates that the optical link could be biased near an SNR-optimal operating point without compromising linearity.  

\subsubsection{MRI measurements}
Although bench measurements quantify the intrinsic gain, noise, and linearity of the analog optical link, these standalone metrics do not directly determine MRI performance. In MRI measurements, the optical link is preceded by the LNA, and the overall noise factor is therefore governed by the cascaded gain and noise contributions of the LNA and optical link according to the Friis relationship \cite{friis1944noise}
\begin{equation}
F_{tot} = F_{LNA} + \frac{F_{opt} - 1}{G_{LNA}}
\label{Friis}
\end{equation}
where $F_{LNA}$ and $G_{LNA}$ are the noise factor and gain of the LNA in linear region, and $F_{opt}$ is the noise factor of the optical link. The relevant metric for MRI is image SNR under realistic coil loading and receiver chain conditions. Figure \ref{fig2}h shows the signal, noise, and image SNR extracted from phantom images in Figure \ref{fig2}g as a function of MZM bias at a fixed optical power of 10\,\unit{\mW} for different LNA gain settings. The image-domain signal (top) followed the bias-dependent optical link gain profile as expected. In contrast, at high LNA gain, the image noise was dominated by the sample and coil noise, therefore followed the signal gain curve more closely. At lower LNA gain settings or near the MZM transfer function extrema, the contribution from the optical link became more prominent, and the image noise approached a floor set by the optical link noise. 

As a consequence, the bias condition that maximized image SNR did not coincide with the quadrature condition. Instead, the SNR curves were asymmetric and resembled an inverted NF profile of the optical link, with the maximum occurring near a bias angle of approximately $\theta$\,=\,$0.2\pi$ for an LNA gain of 48\,\unit{\decibel} (Figure \ref{fig2}h, bottom). At this operating point, the system reached the reference SNR (dashed line). At the highest LNA gain of 48\,\unit{\decibel}, data points near and beyond the quadrature bias point were excluded because the combined LNA and optical link gain saturated the receiver. These results demonstrate that the maximum-gain bias condition does not necessarily maximize image SNR; instead, a lower-bias operating point can improve SNR by reducing link-added noise relative to sample noise.

Figure \ref{fig2}i shows signal, noise, and SNR as a function of input optical power to the MZM at the SNR-optimal bias point identified above. For all LNA gain settings, increasing the input optical power improved image SNR, agreeing with the enhancement of NF at higher input optical power. The SNR improvement was most pronounced at low optical powers and became more gradual at higher optical powers, suggesting a transition from a link-noise-limited regime to one dominated by coil, sample, and LNA-added noise. Increasing the LNA gain also improved SNR and reduced the optical power required to approach the galvanic reference, in line with the Friis relationship (Eq. \ref{Friis}). At an LNA gain of 48\,\unit{\decibel} and $>$5\,\unit{\mW} optical power at the MZM input, the system matched the reference SNR. Subsequent MRI experiments were therefore performed at the optimized low-bias condition and high LNA gain, with 10\,\unit{\mW} optical power providing additional operating margin.

\subsection{Optical power transmission}
Figure \ref{fig3}a illustrates the single-channel Light Coils system with entirely optical data communication, power transfer, and detuning. An 850\,\unit{\nm} laser source was split into two optical arms. One arm was coupled to a GaAs-based PPC, i.e. a photovoltaic cell optimized for conversion of laser light \cite{helmers20206, helmers202168, Algora2022}, which supplied electrical power to two cascaded low-power LNAs. The second arm was routed to the detuning circuit through an MRI-triggered optical switch and a variable optical attenuator, enabling control of coil detuning. This configuration allowed both LNA powering and transmit-phase coil detuning to be driven optically, while keeping unwanted noise sources in active electronic devices outside the MRI room.

\subsubsection{LNA powering with the PPC}
The PPC-powered cascaded LNAs were characterized on the bench and compared with two reference power configurations: a direct DC power supply and a battery-powered voltage regulator. LNAs drew 57.3\,\unit{\mA} at 1.2\,\unit{\V} from both the DC power supply and the regulated battery supply, showing no measurable degradation in NF or gain. For optical powering, a single PPC generated an open-circuit voltage of approximately 1.15\,\unit{\V}, while the available current was controlled by the incident optical power in accordance with its current-voltage (I-V) characteristics (Supplementary Figure S3).

As shown in Figure \ref{fig3}b and c, the LNA NF decreased rapidly with increasing optical power $P_{\mathrm{opt}}$ delivered to the PPC and approached the DC-powered reference once $P_{\mathrm{opt}}$ exceeded 80\,\unit{\mW}. In parallel, the LNA gain increased with optical power and gradually saturated at 50\,\unit{\decibel} (i.e., at high optical powers), approaching the gain obtained in the battery-powered configuration. Across the measured range of $P_{\mathrm{opt}}$, the PPC-powered LNA exhibited slightly lower gain than the electrically powered references due to lower voltage generated by the PPC, but still reached up to 48\,\unit{\decibel} and an NF of 0.93\,\unit{\decibel} when the incident optical power was 100\,\unit{\mW}. Based on the system-level requirements established in the optical data transmission subsection, this gain and noise performance was considered sufficient to implement the optically powered LNA in the optical data transmission system. 

\subsubsection{Optical detuning}
\begin{figure}[ht!]
    \centering
    \includegraphics[width=1\linewidth]{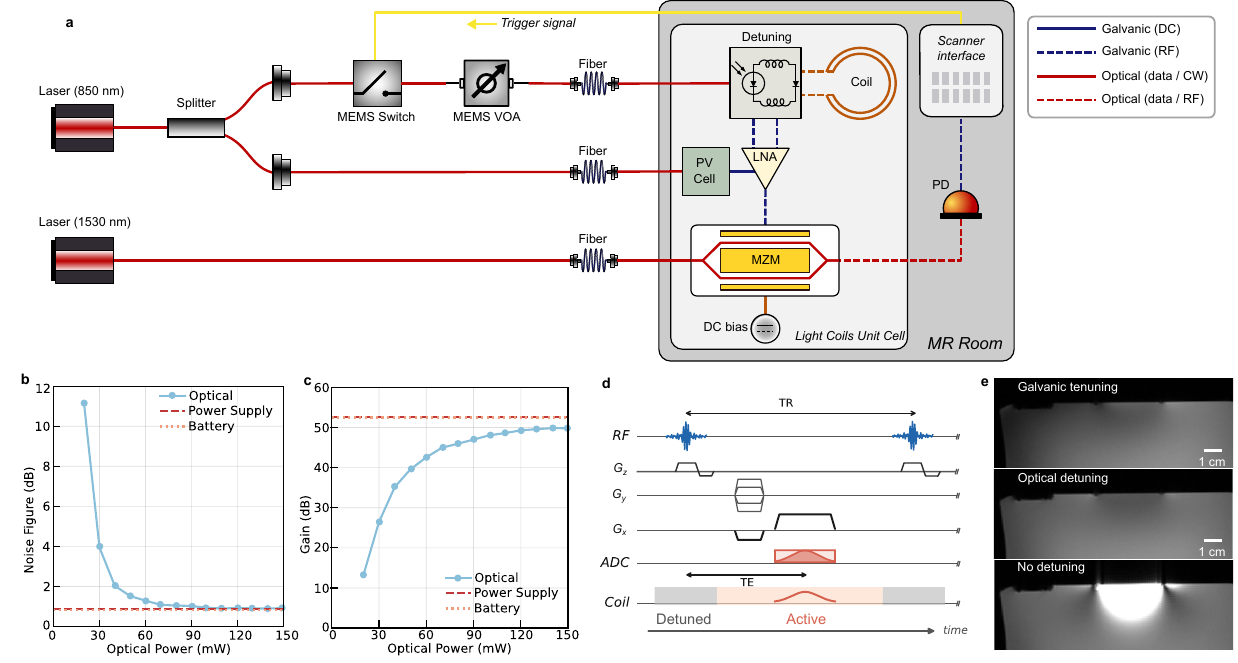}
    \caption{\textbf{Characterization of the optical power transmission. (a)} Schematic of the optical power transmission line integrated with data transmission. Red lines denote optical paths, blue solid lines electrical connections, and dashed lines trigger/control signals. \textbf{(b)} FLASH-based sequence diagram showing the synchronization between the RF excitation, gradient waveforms (G\textsubscript{x,y,z}), ADC window, and the coil state. The coil is held in the detuned state during RF transmission and switched to the active state during the readout. \textbf{(c)} Phantom images acquired with the body coil while the surface coil is detuned by three different mechanisms: conventional galvanic (PIN-diode) detuning (left), 10\,\unit{\mW} optical detuning (middle), and no detuning (right). \textbf{(b)} noise figure and \textbf{(c)} gain of the PPC powered LNA chain as a function of the optical power delivered to the PPC. Dashed red lines correspond to a benchtop DC power supply and dotted orange lines to battery powering, both used as references.}
    \label{fig3}
\end{figure}

Optical detuning was implemented in the MRI sequences (e.g., a spoiled gradient echo 2D FLASH sequence) by incorporating a trigger signal (Figure \ref{fig3}d), so that during RF transmission, the optical switch delivered laser to the detuning circuitry, driving the receive coil into a detuned state; during signal reception, the switch blocked the laser, allowing the coil to return to resonance at the Larmor frequency. The dependence of the optical detuning on laser power was evaluated in phantom experiments using the body coil for both transmission and reception, with the surface coil placed on the phantom. At an optical power of 10\,\unit{\mW}, no signal amplification near the coil was seen so that optical active detuning produced images comparable to those obtained with galvanic detuning, whereas disabling detuning led to a pronounced bright artifact at the coil position due to coupling between the surface coil and the body-coil transmit field (Figure \ref{fig3}e). These results confirm that optical detuning effectively isolates the receive coil during RF transmission while preserving its resonant state during reception.

\subsection{MRI validation of single-channel Light Coil}
\subsubsection{Phantom measurements}
Phantom experiments were performed to evaluate the single-channel Light Coil system (Figure \ref{fig3}a) under different optical power conditions. The power of the laser driving the PPC was varied to control the electrical power available to the cascaded LNAs, and the MZM input optical power was adjusted to tune the optical link performance.

\begin{figure}[ht!]
    \centering
    \includegraphics[width=1\linewidth]{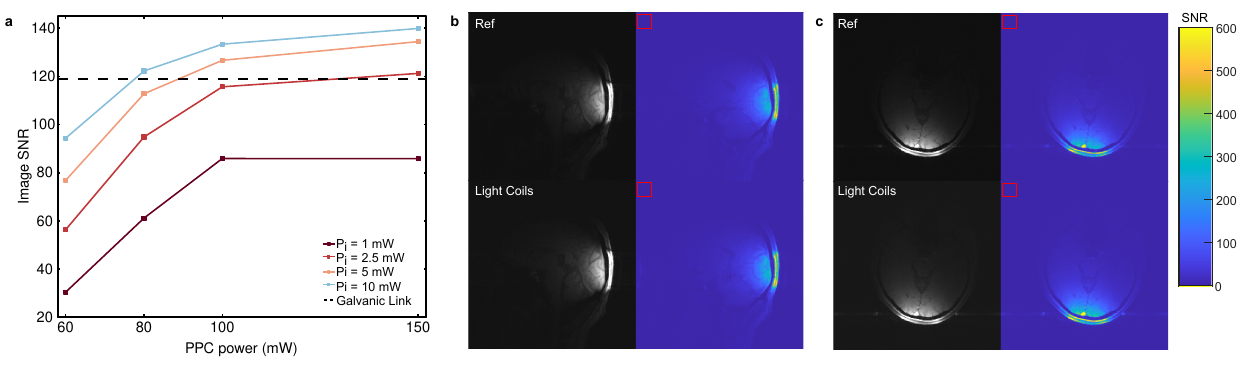}
    \caption{\textbf{MRI measurements of the single-channel Light Coil. (a)} Image SNR measured from the phantom using the sequence in Figure \ref{fig3}b as a function of the PPC power, for four different laser power levels Pi used for the optical data link (1, 2.5, 5, and 10\,\unit{\mW}). The black dashed line indicates the SNR obtained with a conventional galvanic-link reference coil. \textbf{(b)} and \textbf{(c)} are \textit{in vivo} brain images acquired with the same sequence. For two slice orientations (sagittal, (b); transverse, (c)), magnitude images and corresponding SNR maps are shown for the conventional reference coil (top row, "Ref") and the Light Coil (bottom row, "Light Coil"). SNR maps are displayed on a common color scale (0–600 a.u.); red squares indicate the background ROI used for noise estimation.}
    \label{fig4}
\end{figure}

As shown in Figure \ref{fig4}a, increasing the optical power sent to the PPC improved image SNR for all tested MZM input powers. The SNR increased from 30.7 to 37.9 for MZM input powers between of 1 and 10\,\unit{\mW}, as the PPC input power was increased from 60\,\unit{\mW} to 80\,\unit{\mW}, consistent with the LNA entering a low-noise, high-gain regime. Above 100\,\unit{\mW}, further increases in PPC input power yielded only modest SNR gains, indicating that the PPC supplied sufficient electrical power for the LNAs. At a fixed PPC input power, increasing the MZM input power from 1 to 10\,\unit{\mW} also improved image SNR by enhancing the optical link performance, before the LNA reached its optimal operating regime. Because the optical link acts as a later stage in the receiver chain, the benefit of improving its gain or noise performance diminishes according to the Friis formula (Eq. \ref{Friis}), explaining the reduced SNR gain when increasing MZM power from 5 to 10\,\unit{\mW}. At higher optical powers, the optical receiver chain produced SNR values slightly exceeding those of the galvanic reference, which we attribute to its higher overall gain.

Based on these measurements, an SNR comparable to the galvanic link could be achieved with 80\,\unit{\mW} PPC power and 5\,\unit{\mW} MZM input power, corresponding to a total optical power of about 95\,\unit{\mW} per channel, including 10\,\unit{\mW} for optical active detuning. A more conservative operating condition with additional margin was achieved with 100\,\unit{\mW} PPC power, 10\,\unit{\mW} MZM input power and 10\,\unit{\mW} detuning power (120\,\unit{\mW} total), which we used for the subsequent \textit{in vivo} measurements. 

\subsubsection{\textit{In vivo} measurements}
Figure \ref{fig4}b depicts representative human brain images acquired with a single-channel Light Coil and with the galvanic reference in sagittal and transverse orientations, together with corresponding SNR maps. The Light Coil images exhibited anatomical contrast and structural detail that closely matched the galvanic reference, with no additional artifacts or visible image degradation. The SNR maps showed the expected sensitivity profile of a local receive coil, with the highest SNR near the coil and a gradual decrease with distance, and indicated that the Light Coil provided comparable SNR to the galvanic reference, consistent with the phantom results. These findings demonstrate that simultaneous optical data and power transmission, combined with optical detuning, can provide MR image data of comparable quality to that of a conventional galvanic link \textit{in vivo}.

\subsection{Four-channel Light Coil with DWDM}
\begin{figure}[ht!]
    \centering
    \includegraphics[width=1\linewidth]{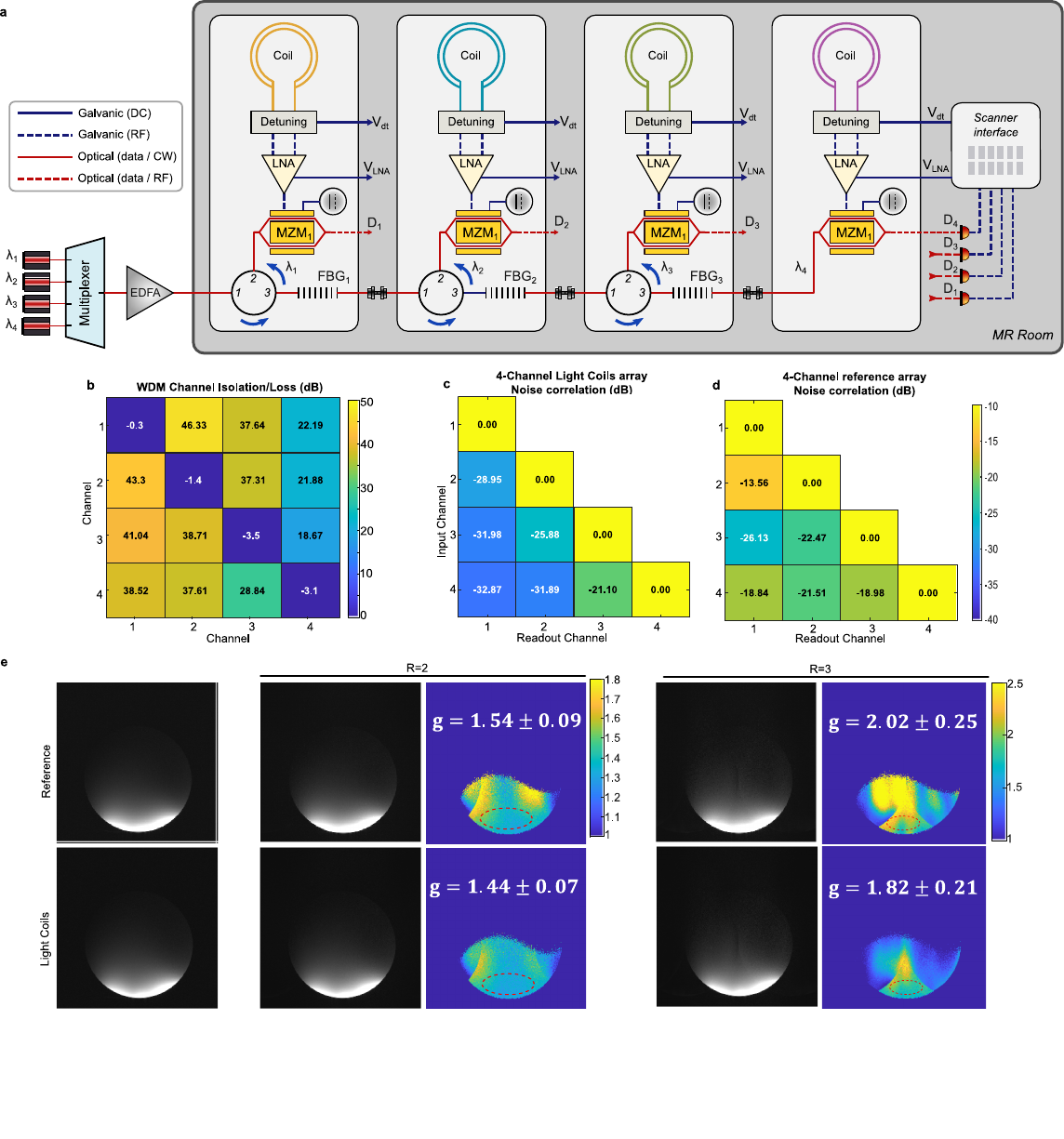}
    \caption{\textbf{Performance 4-channel Light Coils. (a)} Schematic of the 4-channel Light Coil array. \textbf{(b)} Optical channel isolation/loss matrix of the WDM network. Diagonal entries indicate the optical loss of the addressed wavelength–channel, whereas off-diagonal entries (in \unit{\decibel}) quantify the optical leakage between input channels (rows) and output channels (columns). \textbf{(c)} and \textbf{(d)} Noise correlation matrices (in \unit{\decibel}) computed from noise only scan for the 4-channel Light Coil array and a geometrically matched four-channel reference (coax cable) array. The Light Coil array exhibits substantially lower inter-channel noise correlation (off-diagonal values between –21.1\,\unit{\decibel} and –32.9\,\unit{\decibel}) than the reference array (–13.6\,\unit{\decibel} to –26.1\,\unit{\decibel}). \textbf{(e)} Phantom images and corresponding g-factor maps acquired with the reference array (top row) and the Light Coil array (bottom row) at parallel-imaging acceleration factors R = 2 and R = 3. g-factor maps are displayed on a common color scale per acceleration factor.}
    \label{fig5}
\end{figure}

We next extended the optical transmission architecture to a four-channel receiver array using DWDM as shown in Figure \ref{fig5}a. Four continuous-wave (CW) laser sources at distinct C-band wavelengths were combined by an optical multiplexer, amplified by a C-band erbium-doped fiber amplifier (EDFA), and delivered into the MRI room through a single optical fiber. The multiplexed optical carrier was distributed to the four receive channels via optical circulators and wavelength-selective fiber Bragg gratings (FBGs), with each FBG reflecting a single wavelength to its corresponding MZM. The MR signal from each coil was then modulated onto its assigned optical carrier and transmitted back to the MRI system receiver interface. Referring to the conceptual Light Coils system depicted in Figure \ref{fig1}, this implementation uses the circulator/FBG pairs cascaded in series as the on-coil demultiplexer. Each channel used the optimized MZM bias and LNA gain (48\,\unit{\decibel}), which yielded an SNR identical to the galvanic reference to evaluate the four-channel DWDM optical data transmission and parallel imaging performance. By assigning a distinct wavelength to each receive channel, this architecture enabled wavelength-selective channel separation, individual channel control and scalable multi-channel optical transmission, while reducing the number of optical fibers required between the control room and the MRI room.

\subsubsection{WDM isolation}
To evaluate wavelength separation in the WDM network, we measured the inter-channel optical isolation by enabling each laser wavelength in turn and recording the optical power received at all four output channels. The resulting isolation matrix is shown in Figure \ref{fig5}b. The diagonal elements represent the optical loss of the addressed wavelength–channel pairs, whereas off-diagonal elements quantify leakage into non-target channels. For channels 1–3, the isolation exceeded 28\,\unit{\decibel}. Channel 4 showed a lower isolation of 18\,\unit{\decibel}, due to the absence of a terminal FBG unit at this channel, resulting in residual leakage from other channels accumulating at the final output. Together with the wavelength-specific power calibration, the isolation provided by the WDM network was sufficient for channel separation in the four-channel Light Coil array.

\subsubsection{Noise correlation}
Noise scans were acquired with the four-channel Light Coil array and the corresponding galvanic reference array using the identical coils. As shown in Figure \ref{fig5}c, d, the Light Coil array exhibited lower noise correlation coefficients. The optical receiver chain showed off-diagonal correlation levels ranging from -21.1\,\unit{\decibel} to -32.9\,\unit{\decibel}, whereas the galvanic reference showed higher correlations, ranging from -13.6\,\unit{\decibel} to -26.1\,\unit{\decibel}. These results indicate that the DWDM system did not introduce measurable correlated noise between receive channels. In addition, the shared optical components, including the EDFA, circulator, and FBG network, did not produce detectable channel-to-channel crosstalk under the tested imaging conditions. Overall, the noise-correlation measurements show that the four-channel Light Coil array preserved channel independence and, in this implementation, reduced inter-channel noise correlation relative to the galvanic reference array. 

\subsubsection{Phantom MRI with parallel imaging}
MRI experiments on a phantom were performed to validate the four-channel Light Coil array under both fully sampled and accelerated acquisitions using Generalized Autocalibrating Partially Parallel Acquisitions (GRAPPA) \cite{griswold2002generalized}. As shown in Figure \ref{fig5}e, fully sampled phantom images acquired with the Light Coil array and with the galvanic reference exhibited similar signal distributions, indicating that the basic receive-coil sensitivity of the four-channel array was preserved in the optical implementation.  

Parallel imaging performance was further evaluated with acceleration factors of R\,=\,2 and R\,=\,3. For both accelerations, the Light Coil array produced reconstructed images that were visually comparable to those obtained with the galvanic reference, with no additional aliasing artifacts or signal loss. The corresponding g-factor maps showed similar spatial distributions and values between the two configurations, with higher g-factor values at R\,=\,3 than at R\,=\,2, as expected. At higher acceleration (\(R = 3\)), the Light Coil array showed a modest reduction in g-factor in the elliptical ROI (\(g = 1.82 \pm 0.21\) vs.\ \(2.02 \pm 0.25\)), corresponding to approximately 10\% lower noise amplification, while g-factor values at \(R = 2\) remained very similar between arrays (\(1.44 \pm 0.07\) vs.\ \(1.54 \pm 0.09\)). This indicates that the Light Coil array matches and may slightly improve parallel-imaging performance compared with the conventional galvanic array. 

\subsubsection{\textit{In vivo} validation}
\begin{figure}[ht!]
    \centering
    \includegraphics[width=1\linewidth]{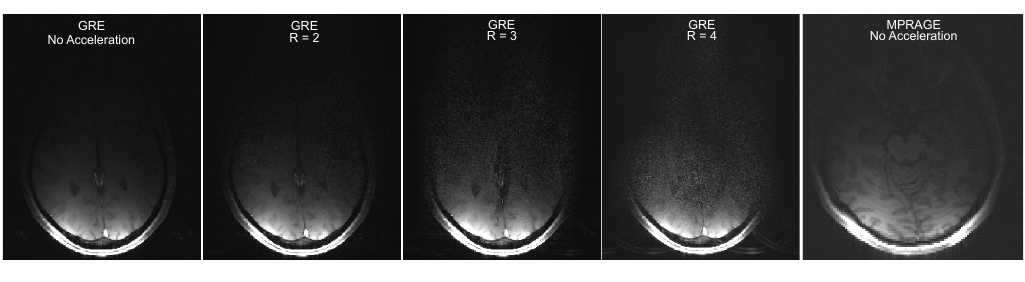}
    \caption{\textbf{\textit{in vivo} brain imaging of 4-channel Light Coil array.} Representative axial images acquired in a healthy volunteer using the proposed 4-channel Light Coil array. Left panels (GRE): 2D-FLASH images obtained without parallel imaging acceleration and with GRAPPA acceleration factors R = 2, R = 3, and R = 4. Right panel (MPRAGE): T1-weighted 3D MPRAGE image acquired without acceleration.}
    \label{fig6}
\end{figure}

Finally, the four-channel Light Coil array was tested for \textit{in vivo} brain imaging in a healthy volunteer. As shown in Figure \ref{fig6}, the fully sampled acquisition of 2D FLASH produced brain images with clear anatomical contrast and the expected local sensitivity profile of the receive array. Parallel imaging acquisitions with acceleration factors of R\,=\,2, 3, and 4 were also reconstructed successfully, demonstrating the compatibility of the Light Coil array with parallel imaging. As expected, increasing the acceleration factor led to higher apparent noise and reduced image quality, consistent with the increased noise amplification observed in the phantom g-factor analysis. Importantly, no additional artifacts attributable to the optical data transmission, DWDM wavelength routing, or shared optical fiber architecture were observed. In addition to 2D FLASH imaging, a T1-weighted MPRAGE dataset was acquired to assess the compatibility of the Light Coil array with 3D MRI. The resulting MPRAGE images showed excellent T1 contrast, clear anatomical structures, and no visible artifacts, indicating that the system can support 3D acquisitions with a higher signal dynamic range. Together, these \textit{in vivo} results show that the four-channel Light Coil supports human brain imaging across accelerated 2D parallel imaging and 3D anatomical protocols.

\section{Discussion}\label{discussion}
In this work, we demonstrated a fully optical receiver chain in MRI, in which data transmission, front-end power delivery, and coil detuning are all implemented optically. The system was first optimized and validated in a single-channel configuration and was subsequently extended to a four-channel receiver array using DWDM for data transmission. Phantom and \textit{in vivo} measurements showed that the optical implementation achieved MRI performance comparable to the galvanic reference while preserving multi-channel scalability and potentially improving parallel imaging capability. These results establish a route towards scalable, dense receive arrays that eliminate galvanic cabling between the coil and the MRI system interface.

A key insight from the optical data transmission experiments is that the optimum operating point of an MZM-based analog optical link should be selected to minimize the effective noise contribution of the link, and not to maximize the RF gain. In a previous implementation of an analog optical link using an MZM \cite{nobre2022optical}, quadrature bias was chosen as it maximizes the small-signal modulation efficiency and minimizes the even-order nonlinear distortions. For MRI signal reception, however, second-harmonic components lie outside the receive band and are further rejected by the receiver filtering. Moreover, the received MR signal, together with coil and sample noise, is first amplified by an LNA prior to the modulator input, and the optical link noise is added after this amplification stage. The final image SNR is therefore determined by the relative contributions of amplified pre-modulator noise and noise added by the optical link. Therefore, the dominant noise source in the cascaded system should be identified and optimized based on the relative noise contributions of the optical components. 

Beyond data transmission, integration of optical power delivery and optical detuning was essential to realize the concept of Light Coil. The SNR dependence on PPC input optical power reflects the load-defined operating point of the PPC-powered LNA. Increasing optical power shifts the solar-cell-like PPC I–V curve and changes its intersection with the LNA load curve. This moves the LNA toward its intended low-noise bias condition, explaining the strong SNR improvement at low PPC powers and the subsequent saturation once sufficient operating current is reached. Therefore, the relevant design criterion is not maximum efficiency alone, but stable delivery of the required LNA current and voltage. Laser intensity noise and intrinsic PPC noise may also cause small supply-current fluctuations, potentially modulating the LNA operating point, and leads a slightly increased NF. Imaging performance comparable to the galvanic reference was achieved with a total optical power of 120\,\unit{\mW} per channel, which is substantially lower than the approximately 370\,\unit{\mW} required by a vendor-supplied conventional galvanically powered and detuned receiver coil (170\,\unit{\mW} for detuning and 200 \,\unit{\mW} for LNA) \cite{aghabagheri2026lightpowered, Gerlach2026_38629167}. This reduced power requirement is particularly important for future dense receive-array implementations, where total power consumption, heat dissipation, and system-integration complexity become increasingly critical. In the present system, optical detuning and PPC-based power delivery were realized at 850\,\unit{\nm} rather than in the C-band, primarily for practical reasons related to component availability and integration. The selected detuning photodiodes provided suitable electrical characteristics and MR compatibility at 850\,\unit{\nm} \cite{gerlach2025opticaldetuning}, while efficient 850\,\unit{\nm} PPCs were readily available for LNA powering \cite{liu2024lightcoils, aghabagheri2026lightpowered}, making them a natural choice for this first Light Coil prototype. Thus, the use of 850\,\unit{\nm} components reflects a pragmatic choice for this prototype rather than a fundamental limitation of the Light Coil architecture, as both detuning and optical powering can in principle be translated to C-band wavelengths.

WDM represents a key step towards extending the Light Coil architecture to dense receive arrays. In our implementation, the combination of optical circulators and wavelength-selective fiber FBGs formed an add–drop-like routing network in which individual receive elements were connected along a shared optical fiber. This modular configuration is well suited for the vision of Light Coils, in which the number and arrangement of coil arrays can be adjusted to different anatomical targets or patient sizes. Independently tunable laser sources provided channel-specific control of wavelength and optical power, enabling per-channel optimization while maintaining a single-fiber connection between the control room and the MRI room. The measured DWDM isolation showed that optical crosstalk between channels could be kept sufficiently low for MRI, although the reduced isolation in the last channel underscored the importance of appropriate optical termination in cascaded FBG networks.

Noise-correlation measurements further confirmed that the DWDM-based optical transmission network preserved channel independence and, in our implementation, reduced noise correlation relative to the galvanic reference. Replacing multi-channel coaxial cables with optical fibers removes galvanic connections between the RF coils and the scanner receive interface, suppressing shared ground paths, common-mode current paths and cable-to-cable coupling that can mediate correlated noise. In addition, noise sources introduced by the optical links, such as wavelength-specific laser intensity noise and photodetector shot noise, are largely uncorrelated between channels and therefore reduce normalized noise-correlation coefficients, even if they do not necessarily increase absolute SNR. The parallel imaging experiments showed that the four-channel Light Coil array preserved the effective coil-sensitivity encoding and did not introduce substantial additional noise amplification, as reflected in g-factor maps lower than the galvanic reference at acceleration factors up to R\,=\,3. Together, these observations indicate that multiplexing and demultiplexing can be incorporated without compromising the independence required for multi-channel MRI.

Several limitations remain in the present proof-of-concept implementation. First, the optical data link was constrained by the optical power-handling capabilities of the MZM and the photodiode. Operation at higher MZM input powers, particularly in the low-bias regime, could further improve the intrinsic noise figure and SFDR\textsubscript{3} of the optical link and thereby reduce the LNA gain needed to preserve system-level SNR. Moreover, the MZM bias was fixed during experiments; practical implementations will require active bias control to compensate for drift during long acquisitions or changing environmental conditions. Second, translating optical power delivery and optical detuning to C-band wavelengths could further improve system integration and reduce the number of distinct optical subsystems. This could be achieved by leveraging S-band InGaAsP or C-band InGaAs PPCs \cite{forcade2025multi, helmers2022unlocking, fafard2022high}. However, their different I-V characteristics would require redesign of the LNA biasing and power-distribution network to maintain the desired gain and NF. Third, the MEMS optical switch used for detuning introduces switching times and delays that may limit compatibility with pulse sequences requiring very short echo times. Faster optical switching technologies would be preferable for broader compatibility across sequences. Finally, the current system uses off-the-shelf optical components that are not specifically designed for MRI environments. Although we did not observe image artifacts attributable to these components, future dense arrays will require MRI-compatible packaging, low-loss interconnects, and higher-power optical components optimized for use inside or near the scanner bore.

Beyond the present implementation, the broader potential of Light Coils lies in overcoming several constraints of conventional galvanic links, thereby enabling lightweight, dense arrays with channel counts that exceed what is practical with traditional cabling. This vision could be supported by integrating optical data transmission and DWDM components into photonic integrated circuits, reducing system size, complexity, and channel-to-channel variability. Combined with modular, flexible coil arrays, such integrated systems could support dense receive arrays whose geometry and channel count can be adapted to different anatomical targets and patient sizes. The broad bandwidth of the optical components also suggests that the approach is compatible with frequencies spanning from ultra-low-field to ultra-high-field MRI, raising the possibility that Light Coil concepts could be translated across a wide range of MRI platforms.

\section{Methods}\label{method}
\subsection{Bench characterization of analog optical link}
A 1550\,\unit{\nm} laser source (TLX1, Thorlabs Inc., Newton, NJ, USA) was coupled to the optical input of an MZM (LN2322, Thorlabs Inc., Newton, NJ, USA). The modulated optical output was detected by a high-speed photodetector (DET01CFC, Thorlabs Inc., Newton, NJ, USA), and the resulting RF signal was measured with a spectrum analyzer (N9000B, Keysight, Germany). A sinusoidal waveform at 123.2\,\unit{\MHz} with an amplitude of -30\,\unit{\decibel m} was generated using an RF signal generator (N5171B, Keysight, Germany) and applied to the RF port of the MZM to emulate the MR signal. The MZM bias voltage was supplied by a DC power supply (HMP4040, Rohde \& Schwarz, Germany) and swept over the full $\textrm{V}_\pi$ range. Measurements were repeated for nominal MZM input optical powers between 1 and 15\,\unit{\mW}. From these data, link gain, noise power spectral density, noise figure, and third-order spurious-free dynamic range were extracted as functions of MZM bias and input optical power.

The small-signal link gain was calculated as the difference between the measured RF output power and the applied RF input power. The measured output noise power spectral density and link gain were subsequently used to calculate the link NF \cite{marpaung_high_2009} with
\begin{equation}
NF_{opt} = P_{N,opt} - G_{opt} + 174
\end{equation}
where $P_{N,opt}$ is the output noise power spectral density in \unit{\decibel\milli\per\hertz}, $G_{opt}$ is the measured link gain in \unit{\decibel}, and -174\,\unit{\decibel\milli\per\hertz} is the thermal noise floor at room temperature 290\,\unit{\K}. The input third-order intercept point of the optical link was calculated as
\begin{equation}
IIP_{3,opt} = \frac{4(V_{\pi,RF})^2}{\pi^2 R}
\end{equation}
Where $\textrm{V}_{\pi,\textrm{RF}}$ is the MZM RF half-wave voltage specified in the manufacturer's datasheet, R\,=\,50\,\unit{\ohm} is the load resistance. Thus, the SFDR\textsubscript{3} of the optical link can be calculated as
\begin{equation}
SFDR_3 = \frac{2}{3}(IIP_{3,opt} - NF_{opt} + 174)
\end{equation}
where $IIP_{3,opt}$ is expressed in \unit{\decibel m}. The resulting SFDR\textsubscript{3} is reported in \unit{\decibel\raiseto{2/3}\hertz}.

\subsection{MRI measurements with the optical link}
All MRI experiments were performed on a 3\unit{\tesla} clinical MRI system (MAGNETOM Cima.X, Siemens Healthineers, Forchheim, Germany). RF transmission was carried out using the integrated body coil, while signal reception was performed using either the Light Coils or the galvanic reference configuration, in which identical coils were used and the same coaxial cable was included between the coil and the optical link. Although in Figure \ref{fig1}, the MR signal is routed outside the magnet room after electro-optic conversion, in this study, the optical MR signal was converted back to an electrical signal inside the magnet and fed to the MRI system receiver to ensure a fair comparison. Shielded loop resonators (SLRs) \cite{ozen2021design,ozen2023scalable} were used as receiver coils and constructed with the same loop geometry and size. Phantom experiments were conducted using a homogeneous spherical phantom (Volume\,=\,1000\,\unit{\ml}, T1\,=\,451$\pm$48\,\unit{\ms}, T2\,=\,95$\pm$12\,\unit{\ms}). Image SNR was calculated using an ROI-based method \cite{dietrich2007measurement,henkelman1985measurement}.

MRI experiments were performed using the same 1550\,\unit{\nm} laser source, MZM, and photodetector as described for bench characterization. The receive coil was connected to the RF input of the MZM via a gain stage consisting of two cascaded LNAs (ZX60-P103LN+, Mini-Circuits, USA) powered by an MR-compatible 12-\unit{\V} battery through a voltage regulator to generate 5\,\unit{\V}. To evaluate the effect of LNA gain on link performance, different gain settings were implemented by cascading one or two LNAs and, where required, inserting fixed RF attenuators (VAT-A-SERIES, Mini-Circuits, USA) before the MZM RF input. For each gain setting, phantom images were acquired while varying the MZM input optical power and sweeping the MZM bias voltage over one full $\textrm{V}_\pi$ range in a step of 0.2\,\unit{\V}. In reference measurements using galvanic connections, the same receive coil was connected directly to the MRI system’s receiver interface using a coaxial cable. For both configurations, coil detuning was performed galvanically via a DC cable connected to the receive interface, and a 2D-FLASH sequence was used with TE/TR\,=\,5/15\,\unit{\ms}, flip angle\,=\,15\,\unit{\degree}, FOV\,=\,256$\times$256\,\unit{\square\meter}, $\Delta\textrm{V}=1\times 1 \times 3$\,\unit{\cubic\mm}. The sequence was programmed in the Pulseq environment \cite{chen2025automated, layton2017pulseq}, and the raw data were exported to an external PC and reconstructed using an FFT algorithm provided by MATLAB (2024, MathWorks, Natick, MA, USA).

\subsection{Optical detuning}
Optical detuning was implemented by connecting a photodiode–PIN diode pair between the inner and outer conductors of the SLR \cite{gerlach2025opticaldetuning}. Details of the detuning circuit are provided in Supplementary Figure S5. The photodiode was coupled to a multimode fiber using a 3D-printed fiber coupler. The detuning performance was first characterized on the bench using a network analyzer (P5050B, Keysight, Germany). The S parameters of the coil were measured without optical illumination to determine the tuned state. An 850\,\unit{\nm} laser source (RTMDL-852-1W, Roithner LaserTechnik, Austria) was then coupled to the multimode fiber through a variable optical attenuator, and the same S parameters were measured at optical powers of 5\,\unit{\mW} and 20\,\unit{\mW} delivered to the photodiode. 

Optical detuning was further evaluated in phantom MRI experiments. The body coil was used for both RF transmission and signal reception, while the surface SLR was placed on top of a bottle phantom (3.75\,g NiSO$_4$ $\times$ 6H$_2$O, 5\,g NaCl per 1000\,g H$_2$O) and was not connected to the scanner receive interface. Images were acquired and compared for conventional galvanic detuning, optical detuning with 10\,\unit{\mW} of constant optical power delivered to the detuning circuit, and no detuning. The same 2D FLASH sequence with TE/TR\,=\,10/100\,\unit{\ms}, flip angle\,=\,25\,\unit{\degree}, FOV\,=\,250$\times$250\,\unit{\square{\mm}}, and $\Delta V$\,=\,1$\times$1$\times$5\,\unit{\cubic\mm} was used.

\subsection{PPC powered LNA}
Optical powering was implemented using a GaAs PPC described in \cite{hohn2016optimal, fakidis2020simultaneous}. A \textit{pn}-homojunction with 3.6\,\unit{\um} GaAs absorber thickness is cladded by higher bandgap passivation layers. On the front side, lateral conduction is facilitated by a highly doped 400\,\unit{nm} AlGaAs window layer \cite{oliva2008gaas}. 1.6$\times$1.6\,\unit{\square{\mm}}-sized chips with a 1\,\unit{\mm} active aperture were realized using standard semiconductor processing (photolithography, metallization, anti-reflection coating, mesa etching, dicing) and mounted on TO-46 sub-mounts for testing. 

The current–voltage characteristics and temporal stability of the PPC output are provided in Supplementary Figure S3. An 850\,\unit{\nm} laser was delivered to the PPC via a multimode fiber, which was coupled to the PPC using a 3D-printed fiber coupler. The PPC output powered both the custom low-power LNA and the commercial LNA stage in parallel. The custom LNA was specifically developed to match the current–voltage characteristics of the PPC and was designed as a two-stage cascaded amplifier, consisting of two transistor stages connected in series, with the RF output of the first stage driving the RF input of the second stage \cite{aghabagheri2026lightpowered}. The RF output of the custom LNA was then fed into a commercial LNA chip with a dedicated PCB. This combination of LNAs was used in subsequent optical powering and MRI experiments. The performance of the PPC-powered LNAs was characterized at 123.2\,\unit{\MHz} using the NF measurement mode of the spectrum analyzer and the calibrated noise source (N4001A, Keysight, USA). The optical power delivered to the PPC-coupled fiber was varied using a variable optical attenuator (VOAMMF, Thorlabs Inc., Newton, NJ, USA). 

\subsection{Single-channel Light Coil MRI validation}
To validate the integrated Light Coil system, we combined the optimized MZM operating condition, PPC-based LNA powering, and sequence-synchronized optical detuning in single-channel MRI experiments. In this configuration, MR signals received by the coil were amplified by the optically powered cascaded LNA and transmitted through the analog optical link, while coil detuning during RF transmission was controlled by the optical switch (MEMS-1X2-M-85-M2-5-09-05-FP, Fibermart, KowLoon, Hongkong). Phantom measurements were performed while varying the PPC optical power (60, 80, 100, 150\,\unit{\mW}) and MZM input power (1, 2.5, 5, 10\,\unit{\mW}) to map image SNR as a function of optical operating conditions. For each condition, phantom images were acquired using a 2D-FLASH sequence with TE/TR\,=\,7/15\,\unit{\ms}, flip angle\,=\,15\,\unit{\degree}, FOV\,=\,250$\times$250\,\unit{\square{\mm}}, $\Delta V$\,=\,1$\times$1$\times$3\,\unit{\cubic\mm}. \textit{In vivo} brain images were acquired at a representative operating point (PPC optical power of 100\,\unit{\mW}, MZM optical power of 10\,\unit{\mW} and, optical detuning power of 10\,\unit{\mW}) to assess anatomical image quality and SNR relative to the galvanic configuration using a 2D-FLASH sequence with TE/TR\,=\,10/150 ms, flip angle\,=\,60\,\unit{\degree}, FOV\,=\,250$\times$250\,\unit{\square{\mm}}, $\Delta V$\,=\,1$\times$1$\times$3\,\unit{\cubic\mm}.

\subsection{Four-channel Light Coil with WDM}
A four-channel coil array was constructed using four geometrically decoupled SLR elements aligned in a row (Supplementary Figure S6\,a). The reflection and transmission coefficients of the array were characterized using the network analyzer (Supplementary Figure S6\,b). Four laser sources (PRO8 DWDM DFB Laser, Thorlabs Inc., Newton, NJ, USA), a multiplexer (FMU-D402160M, FS.com, Germany), and an EDFA (EDFA300S, Thorlabs Inc., Newton, NJ, USA) were placed outside of the MR room. The optical circulators (Agiltron Inc., MA, USA) and FBGs (Technica Optical Components, GA, USA) were mounted on the patient table and together act as a demultiplexer, assigning a separate wavelength to each channel. The output power of each FBG and the fourth circulator was measured to determine the wavelength-dependent optical losses introduced by the circulator, FBGs, fiber couplers, and interfaces. Because the actual reflection peaks of the FBGs deviated slightly from their nominal labeled wavelengths due to fabrication tolerances, the wavelength of each laser source was fine-tuned to match the corresponding FBG reflection peak and maximize the reflected power. Optical loss was compensated by adjusting the output powers of the individual lasers and the EDFA gain, ensuring that comparable optical power was delivered to each MZM, providing a consistent operating condition across the four receive channels.

Phantom measurements with the 4-channel Light Coil were performed using a 2D-FLASH sequence with TE/TR\,=\,5/12\,\unit{\ms}, flip angle\,=\,15\,\unit{\degree}, FOV\,=\,250$\times$250\,\unit{\square{\mm}}, $\Delta V$\,=\,1$\times$1$\times$3\,\unit{\cubic\mm}. Parallel imaging measurements were also performed using GRAPPA with acceleration factors of R\,=\,2 and R\,=\,3. For each accelerated acquisition, 24 auto-calibration signal lines were acquired. Images were reconstructed offline and combined across the receive channels using noise-informed adaptive channel combination. Parallel-imaging noise amplification was calculated using the pseudo-multiple-replica g-factor formulation as described in ~\cite{Robson2008SNRgFactor}, consistent with the GRAPPA g-factor formulation ~\cite{Breuer2009GRAPPAgFactor}. The noise correlation between the channels was measured using a noise-only acquisition with the same 2D FLASH sequence, with the RF transmit voltage set to zero. 

For \textit{in vivo} brain imaging, the coil array was positioned underneath the volunteer’s head over the occipital lobe. Images were acquired using the same 2D FLASH sequence with TE/TR\,=\,5/150\,\unit{\ms}, and flip angle\,=\,60\,\unit{\degree}, to obtain a better T1 contrast. Parallel imaging was also performed with the same acceleration settings as in the phantom measurements. In addition, a T1-weighted MPRAGE image was acquired with TE/TR = 2.6/120\,\unit{\ms}, flip angle\,=\,8\,\unit{\degree}, FOV\,=\,250$\times$250\,\unit{\square{\mm}}, $\Delta V$\,=\,1$\times$1$\times$1\,\unit{\cubic\mm} to demonstrate compatibility with a commonly used \textit{in vivo} brain imaging protocol.

\backmatter

\bmhead{Supplementary information}
Additional details, methods, and results are provided in the Supplementary Information accompanying this manuscript.

\bmhead{Acknowledgements}
The last author gratefully acknowledges the support from Prof. Dr. Maxim Zaitsev and Mojtaba Shafiekhani from the University Medical Center Freiburg in implementing MRI sequences in Pulseq; and the fruitful discussions with Prof. Dr. Jürgen Hennig from University Medical Center Freiburg and Prof. Dr. Ergin Atalar from Bilkent University.

\section*{Declarations}

\begin{itemize}
\item Funding: The work is funded by the German Research Foundation (Deutsche Forschungsgemeinschaft, DFG) – project no: 532643102, awarded to principal investigators ACÖ and ÇA.

\item Conflict of interest/Competing interests: MB and ACO declare joint inventorship of the patent US12292491B2, which covers the fundamental Light Coils concept. Other authors do not declare any conflict of interest.

\item Ethics approval and consent to participate: All human imaging was performed in accordance with relevant guidelines and regulations. Healthy volunteer scanning was approved by the Institutional Review Board of the University Medical Center Freiburg (No. 160/2000), and informed written consent was obtained prior to imaging.

\item Consent for publication: Not applicable.

\item Data availability: The datasets generated and/or analyzed during the current study are not publicly available but are available from the corresponding author upon reasonable request. 

\item Materials availability: Not applicable.

\item Code availability: Not applicable.

\item Author contribution: Z.L. and M.T. designed the experiments, developed the prototypes, performed all measurements, and carried out the primary data analysis. J.G. developed and implemented the optical detuning system and contributed to related measurements, while R.A. designed and characterized the low-power LNA chain and power hub and assisted with measurements. H.H. invented, designed and fabricated the high-efficiency photonic power converters used in this work. M.B. and A.C.Ö. invented the Light Coils concept. M.B. provided conceptual input and critical suggestions during the project. Ç.A. and A.C.Ö. jointly supervised the project and student contributors; they coordinated the experimental and methodological work and data interpretation, with Ç.A. leading the design and integration of the optical setup and components and A.C.Ö. contributing to concept development, experiment design, and system integration. Z.L., M.T., Ç.A. and A.C.Ö. led the manuscript drafting and figure preparation. All authors discussed the results and reviewed and edited the manuscript.

\end{itemize}

\bibliography{sn-bibliography}

\newpage

\newpage

\newpage

\newpage

\newpage

%
%
%
\end{document}


\title[Article Title]{Supplementary Information 

LightCoils: MRI with Fully Optical Power and Data Transmission}

\author*[1]{\fnm{Zining} \sur{Liu}}\email{zining.liu@uniklinik-freiburg.de}
\equalcont{These authors contributed equally to this work.}

\author[2]{\fnm{Morteza} \sur{Teymoori}}\email{morteza.teymoori@imtek.uni-freiburg.de}
\equalcont{These authors contributed equally to this work.}

\author[1]{\fnm{Jakob} \sur{Gerlach}}\email{jakob.gerlach@uniklinik-freiburg.de}

\author[1]{\fnm{Reza} \sur{Aghabagheri}}\email{reza.aghabagheri@uniklinik-freiburg.de}

\author[3]{\fnm{Henning} \sur{Helmers}}\email{henning.helmers@ise.fraunhofer.de}

\author[1]{\fnm{Michael} \sur{Bock}}\email{michael.bock@uniklinik-freiburg.de}

\author[2]{\fnm{Çağlar} \sur{Ataman}}\email{caglar.ataman@imtek.uni-freiburg.de}

\author[1]{\fnm{Ali Caglar} \sur{Özen}}\email{ali.oezen@uniklinik-freiburg.de}

\affil[1]{\orgdiv{Department of Diagnostic and Interventional Radiology}, \orgname{University of Freiburg}, \orgaddress{\street{Killianstr. 5a}, \city{Freiburg}, \postcode{79110}, \state{Baden-Württemberg}, \country{Germany}}}

\affil[2]{\orgdiv{Department of Microsystems Engineering}, \orgname{University of Freiburg}, \orgaddress{\street{Georges Köhler-Allee 102}, \city{Freiburg}, \postcode{79110}, \state{Baden-Württemberg}, \country{Germany}}}

\affil[3]{\orgdiv{Department III-V Photovoltaics and Concentrator Technology}, \orgname{Fraunhofer Institute for Solar Energy Systems ISE}, \orgaddress{\street{Heidenhofstr. 2}, \city{Freiburg}, \postcode{79110}, \state{Baden-Württemberg}, \country{Germany}}}

\maketitle
\newpage
\section{Symbols}
 
\begin{table}[h!]
\centering
\caption{Symbols used in Supplementary Note 1.}
\begin{tabular}{lll}
\toprule
Symbol & Meaning \\ 
\midrule
$\thb$ & MZM bias angle, measured from transmission null \\
$\Vpi$ & MZM RF half-wave voltage at \SI{123.2}{\mega\hertz} \\
$P_i$ & optical power at MZM input \\
$L$ & total optical insertion loss of the link \\
$\mathcal{R}$ & photodiode responsivity at \SI{1550}{\nano\meter} \\
$\Ip$ & photocurrent at maximum transmission, $\frac{\mathcal{R} P_i}{L}$ \\
$\ID$ & average photocurrent, Eq.~\eqref{eq:idc} \\
$\RIN$ & laser relative intensity noise \\
$G_L$, $F$ & on-coil LNA gain and noise factor \\
$R$ & system impedance \\
\bottomrule
\end{tabular}
\end{table}

 
 
 
\section{Transfer function and bias convention}
 
The optical power at the MZM output is
\begin{equation}
P_\mathrm{o}(t) = \frac{P_i}{2L}\Big[1 - \cos\big(\thb + \varphi(t)\big)\Big],
\qquad \thb=\frac{\pi\, v_{DC}}{{\Vpi}_{DC}},
\qquad \varphi(t) = \frac{\pi\, v(t)}{{\Vpi}_{RF}},
\label{eq:transfer}
\end{equation}
where $L$ is the total optical insertion loss of the link (MZM insertion loss, connectors, fibre), $v(t)$ is the RF voltage at the MZM's RF port, and $\thb$ is the bias angle. We measure $\thb$ from the transmission null: $\thb = 0$ corresponds to minimum transmission, $\thb$\,=\,$\pi/2$ to quadrature, and $\thb$\,=\,$\pi$ to maximum transmission. ``Below quadrature'' therefore denotes $0 < \thb < \pi/2$.
 
The average (DC) photocurrent at the detector is
\begin{equation}
\ID(\thb) = \frac{\Ip}{2}\big(1-\cos\thb\big),
\qquad \Ip \equiv \frac{\mathcal{R}\, P_i}{L} ,
\label{eq:idc}
\end{equation}
where $\Ip$ is the photocurrent that would be produced at maximum transmission. Because $\ID$ is a monotonic, single-valued function of $\thb$ over $[0, \pi]$, it serves as the feedback variable for bias control.
 
\section{Small-signal gain}
 
For MR signal levels ($|\varphi| \ll 1$), expanding Eq.~\eqref{eq:transfer} gives the photocurrent
\begin{equation}
i(t) = \frac{\Ip}{2}\left[(1-\cos\thb) + \sin\thb\,\varphi + \frac{\cos\thb}{2}\,\varphi^2 - \frac{\sin\thb}{6}\,\varphi^3 + \mathcal{O}(\varphi^4)\right].
\label{eq:expansion}
\end{equation}
The first-order term yields the small-signal RF power gain of the link (source and load impedance $R$\,=\,50\,\unit{\ohm}, impedance-matched electrode),
\begin{equation}
g(\thb) = \left(\frac{\pi \Ip R}{2 \Vpi}\right)^{2} \sin^2\thb
\;=\; g_q \sin^2\thb ,
\label{eq:gain}
\end{equation}
where $g_q$ is the gain at quadrature. Two scalings follow directly: $g \propto \sin^2\thb$ in bias, and $g \propto P_i^2$ in optical power (through $\Ip^2 \propto P_i^2$).

To extract the link parameters, we measured the static transfer curve of the MZM (optical power after the modulator, recorded with an optical power meter as a function of bias voltage) and the RF gain profile of the complete link. Fitting the transfer curve to Eq.~\eqref{eq:transfer} and the gain profile to Eq.~\eqref{eq:gain} yields the parameters summarised in Table~\ref{tab:fit}.
 
\begin{table}[h!]
\centering
\caption{Optical link parameters extracted from the measured MZM transfer curve (Eq.~\eqref{eq:transfer}) and the measured RF gain profile (Eq.~\eqref{eq:gain}).}
\label{tab:fit}
\begin{tabular}{llll}
\toprule
Parameter & Symbol & Fitted value & Source \\
\midrule
Half-wave voltage (bias electrode) & ${\Vpi}_\mathrm{DC}$ & 5.396\,\unit{\V} & transfer curve \\
Extinction ratio & & 21.7 \,\unit{\dB} & transfer curve \\
Half-wave voltage (RF electrode, 123.2\,\unit{\mega\hertz}) & ${\Vpi}_\mathrm{RF}$ & 5.409\,\unit{\V} & gain profile \\
Link gain/$P_i$ (\unit{\dB m}) slope & $g_q - P_i$ & 2.041 & gain profile \\
Laser RIN & RIN & -163.0\,\unit{\dB\per\hertz}& Noise measurement\\
\bottomrule
\end{tabular}
\end{table}

\section{Noise model and optimum bias}
 
For completeness, we restate here the three noise power spectral densities (in $\mathrm{A^2\,Hz^{-1}}$) added by the link at the photodetector output, which are also given in the main text.
\begin{equation}
\begin{aligned}
N_\mathrm{th} &= \frac{4 k_B T}{R}, \\
N_\mathrm{sh} &= 2 q I_D = q I_P\,(1-\cos\theta), \\
N_\mathrm{RIN} &= \mathrm{RIN}\cdot I_D^{2}
= \frac{\mathrm{RIN}\, I_P^{2}}{4}(1-\cos\theta)^{2}.
\end{aligned}
\label{eq:noisesources}
\end{equation}
where $q$ the elementary charge, $\kB$ the Boltzmann constant and $T$\,=\,290\,\unit{\kelvin}. The thermal term is bias-independent; the shot-noise term scales linearly and the RIN term quadratically with the DC photocurrent.
 
In the full receive chain, the dominant noise at the link input is the LNA output noise, i.e., amplified sample noise plus the LNA's added noise, with density $N_\mathrm{in} = G_L F \kB T$ at the MZM RF port. This term is co-amplified with the signal by $g(\thb)$ and therefore does not by itself shift the optimum bias; it does, however, determine how much reducing the latter actually improves the SNR. Referring the link-added noise to the link input,
\begin{equation}
N_\mathrm{add}(\thb) = \frac{N_\mathrm{sh} + N_\mathrm{RIN} + N_\mathrm{th}}{g(\thb)} ,
\label{eq:nadd}
\end{equation}
and writing $s \equiv 1 - \cos\thb$ ($s = 1$ at quadrature) with $\sin^2\thb = s(2 - s)$,
\begin{equation}
N_\mathrm{add}(s) = \frac{a\,s + b\,s^{2} + c}{g_q\, s\,(2-s)},
\qquad
a = q \Ip,\quad b = \frac{\RIN\,\Ip^{2}}{4},\quad c = \frac{4 \kB T }{R}.
\label{eq:nadds}
\end{equation}
The three contributions scale differently with bias:
\begin{equation}
\frac{N_\mathrm{sh}}{g} \propto \frac{1}{1+\cos\thb},
\qquad
\frac{N_\mathrm{RIN}}{g} \propto \tan^{2}\!\frac{\thb}{2},
\qquad
\frac{N_\mathrm{th}}{g} \propto \frac{1}{\sin^{2}\thb}.
\label{eq:scalings}
\end{equation}
Moving the bias from quadrature towards the null, the input-referred RIN contribution falls steeply and the shot-noise contribution falls moderately, because the optical carrier power --- and with it the detected DC photocurrent --- drops faster than the link gain. The input-referred thermal contribution, in contrast, is minimised exactly at quadrature and grows towards the null as the gain collapses. The competition between these trends places the noise-optimal bias below quadrature whenever shot and RIN noise are non-negligible relative to the receiver thermal floor.
 
Setting $\mathrm{d}N_\mathrm{add}/\mathrm{d}s = 0$ in Eq.~\eqref{eq:nadds} yields a closed-form optimum,
\begin{equation}
s^{*} = \frac{\sqrt{c^{2} + 2c\,(a + 2b)} - c}{a + 2b},
\qquad
\cos\thb^{*} = 1 - s^{*},
\label{eq:optimum}
\end{equation}
which reduces to $s^{*} = 1$ (quadrature) in the thermal-noise limit ($a, b \to 0$) and moves monotonically toward the null as the optical power --- and hence $a$ and $b$ --- increases. For the parameters of our link ($\mathrm{RIN}$ \,=\,-163\,\unit{\dB\per\hertz}), Eq.~\eqref{eq:optimum} predicts $\thb^{*}$ (Fig S\ref{fig:S2}), which is consistent with the bias sweep in Fig. 2, where the measured NF minimum shifts towards lower bias points as optical power increases.
 
Because the LNA output noise $N_\mathrm{in}$ dominates the input-referred total, the achievable SNR improvement is bounded: the system SNR is proportional to $\left[N_\mathrm{in} + N_\mathrm{add}(\thb)\right]^{-1}$, so reducing $N_\mathrm{add}$ from its quadrature value to its minimum improves the noise figure less substantially for our parameters, explaining the $\approx 5\,\%$ improvement observed in imaging experiments. The same expression shows why operating an analogue link without an on-coil LNA would make the bias optimisation far more consequential, at the cost of an unacceptable overall noise figure.
 
The link noise figure in isolation (50,\unit{\ohm} source at $T$\,=\,290\,\unit{\kelvin}, no LNA) follows from Eq.~\eqref{eq:nadd} as $F_\mathrm{link}(\thb) = 1 + N_\mathrm{add}(\thb)/(\kB T)$ and is reported in Fig. 2d as a function of bias and optical power.
 
\section{Linearity}
\label{sec:linearity}
 
The quadratic and cubic terms of Eq.~\eqref{eq:expansion} determine the link's distortion. For a two-tone input $v(t) = V_0(\cos\omega_1 t + \cos\omega_2 t)$, the third-order intermodulation products at $2\omega_1 - \omega_2$ and $2\omega_2 - \omega_1$ have the amplitude ratio to the fundamental of $(\pi V_0/\Vpi)^2/8$, independent of bias, because both the fundamental and the IMD3 terms carry the same $\sin\thb$ prefactor. The input-referred third-order intercept is therefore bias-independent,
\begin{equation}
\mathrm{IIP3} = \frac{4 \Vpi^{2}}{\pi^{2} R},
\label{eq:iip3}
\end{equation}
while the output intercept scales with the gain, $\mathrm{OIP3}(\thb) = g(\thb)\cdot\mathrm{IIP3} = \Ip^2 R \sin^2\thb$. At quadrature this reduces to the familiar $\mathrm{OIP3} = 4 \ID^2 R$. The third-order-limited spurious-free dynamic range is
\begin{equation}
\mathrm{SFDR}_3 = \left[\frac{\mathrm{OIP3}}{N_\mathrm{out}(\thb)}\right]^{2/3}
= \left[\frac{\mathrm{IIP3}}{N_\mathrm{add}(\thb)}\right]^{2/3}
\quad (\dBHz),
\label{eq:sfdr}
\end{equation}
with $N_\mathrm{out} = N_\mathrm{sh} + N_\mathrm{RIN} + N_\mathrm{th}$ the total link-added output noise density. Since IIP3 is bias-independent, $\mathrm{SFDR}_3$ is maximised at the same below-quadrature bias that minimises $N_\mathrm{add}$. Measured $\mathrm{SFDR}_3$ at 123.2\,\unit{\mega\hertz} for different power levels and various bias points is reported in Fig2\,e.
 
Second-order distortion, governed by the $(\cos\thb/2)\,\varphi^2$ term, vanishes at quadrature and grows as the bias moves towards the null. For the narrowband MRI receive signal centred at 123.2\,\unit{\mega\hertz}, the resulting second harmonic ($\approx$ 246\,\unit{\mega\hertz}) and the difference-frequency products (near DC) fall far outside the receive bandwidth and are rejected by the scanner receiver; second-order distortion therefore does not constrain the choice of bias in this application. This is a notable relaxation relative to broadband RF-over-fibre links, where sub-quadrature biasing must trade even-order distortion against noise.

\newpage
\FloatBarrier
\section{Supplementary Figures and Tables}
\begin{figure}[ht!]
    \centering
    \includegraphics[width=1\linewidth]{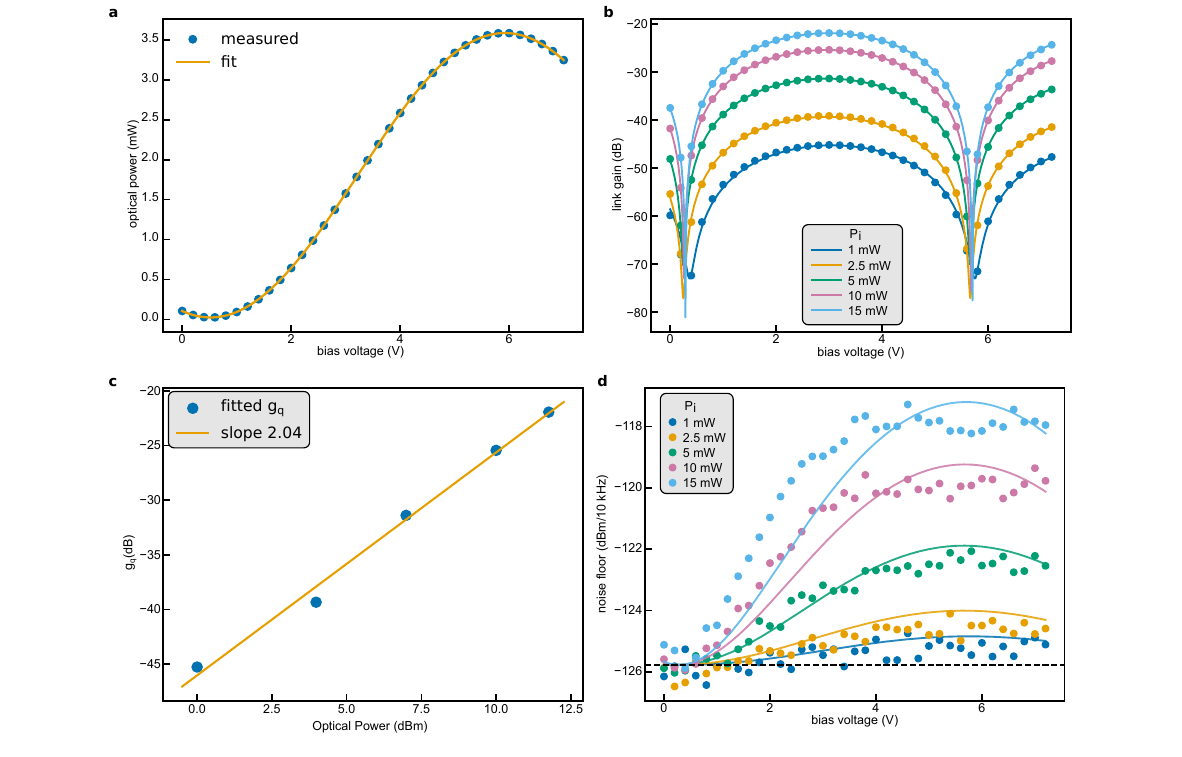}
    \caption{\textbf{Bench characterisation of the analogue optical link.} \textbf{a}~Transfer curve of the Mach--Zehnder modulator: optical power after the modulator versus bias voltage at an input optical power of $P_i$\,=\,10\unit{\milli\watt} (one of five measured sweep cycles shown; circles, power-meter measurement; line, fit to Eq.~\eqref{eq:transfer}). The fit yields ${\Vpi}_\mathrm{DC}$\,=\,5.396\,\unit{\volt}, a total optical loss of $L$\,=\,2.8(-4.5\,\unit{\decibel}) and an extinction ratio of 21.7\,\unit{\decibel}. Across the five consecutive cycles the null position drifts by 49\,\unit{\milli\volt}, while ${\Vpi}_\mathrm{DC}$ remains stable to within 3\,\unit{\milli\volt}. \textbf{b}~RF link gain versus bias voltage for input optical powers of 1-15\,\unit{\milli\watt}, measured with a single tone at 123.2\,\unit{\mega\hertz} (-30\,\unit{\dB m}) on a spectrum analyser (circles, measured tone power referred to the input; lines, curve fits of $g_q \sin^2\!\big(\pi(V-V_0')/V_\pi'\big)$ per dataset). The two gain nulls correspond to the transmission minimum ($\approx$\,0.3\,\unit{\volt}) and maximum ($\approx$\,5.7\,\unit{\volt}) of the modulator; the fitted $V_\pi'$\,=\,5.41$\pm$0.01\,\unit{\volt} at 123.2\,\unit{\mega\hertz} agrees with ${\Vpi}_\mathrm{DC}$ from \textbf{a}. \textbf{c}~Quadrature link gain $g_q$ extracted from the fits in \textbf{b} versus input optical power. The fitted slope of 2.04 confirms the predicted $g_q \propto P_i^{2}$ scaling (Eq.~\eqref{eq:gain}). \textbf{d}~Noise floor of the spectrum-analyser traces (10\,\unit{\kilo\hertz} resolution bandwidth) versus bias voltage for the same datasets (circles). At low optical power the floor coincides with the instrument noise level (dashed line, -165.8\,\unit{\dB m\per\hertz}); at higher power it rises towards the transmission maximum, following the detected-power dependence of shot and laser intensity noise. The fit of experimental datasets yields a laser relative intensity noise of -163.0$\pm$0.6\,\unit{\decibel\per\hertz} at 123.2\,\unit{\mega\hertz}, together with the instrument floor.}
    \label{fig:S1}
\end{figure}

\begin{figure}[h]
    \centering
    \includegraphics[width=1\linewidth]{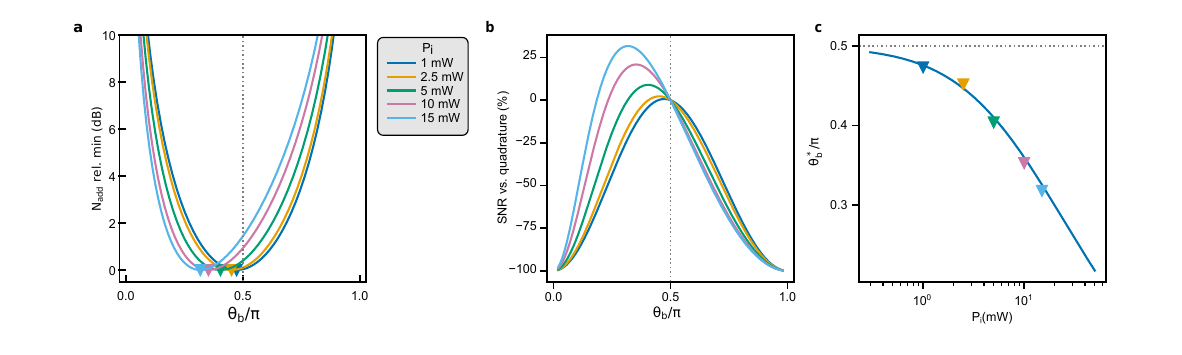}
    \caption{\textbf{Model predictions for noise-optimal biasing of the analogue optical link.} All curves use the link parameters fitted in Supplementary Fig. S~\ref{fig:S1} ($g_q$, $V_\pi'$, and the measured laser relative intensity noise of -163.0\,\unit{\decibel\per\hertz}); the bias angle $\thb$ is measured from the transmission null, with quadrature at $\thb = \pi/2$ (dotted lines). \textbf{a}~Link-added noise referred to the link input, $N_\mathrm{add}(\thb)$ (Eq.~\eqref{eq:nadds}), relative to its minimum, for input optical powers of 1-15\,\unit{\milli\watt}. Triangles mark the closed-form optimum $\thb^{*}$ (Eq.~\eqref{eq:optimum}). At low optical power the thermal noise of the 50\,\unit{\ohm} termination dominates and the optimum coincides with quadrature; with increasing power the growing shot and intensity-noise contributions, which fall faster than the link gain towards the null, shift the optimum below quadrature. \textbf{b}~Predicted system SNR versus bias, relative to quadrature operation, including the LNA-dominated noise at the link input. The input noise term bounds the achievable improvement; the position of the maximum is independent of $G_L$ and $F_L$ and coincides with $\thb^{*}$ from \textbf{a}. \textbf{c}~Optimum bias angle $\thb^{*}$ versus input optical power (line, closed-form solution of Eq.~\eqref{eq:optimum} using the fitted link transmission and intensity noise; triangles, values for the five measured datasets, colours as in \textbf{a}). Because the DC bias point drifts between experiments, the optimum is specified as a bias angle --- equivalently a detected-power set point for the bias controller --- rather than an absolute voltage.}
    \label{fig:S2}
\end{figure}

\begin{figure}[h]
    \centering
    \includegraphics[width=1\linewidth]{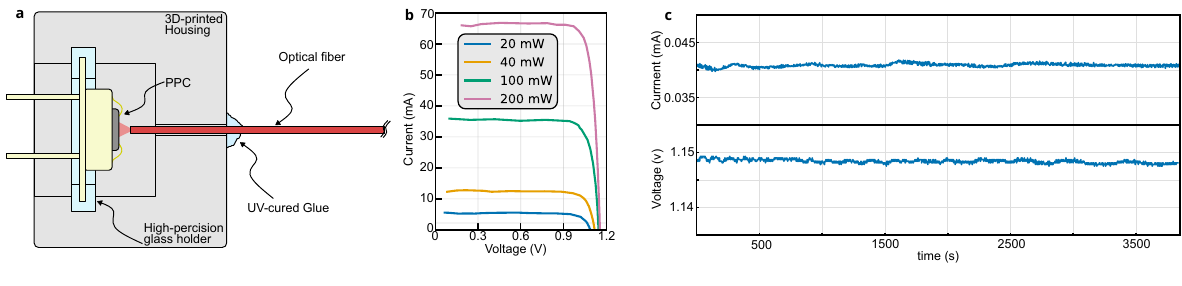}
    \caption{\textbf{Photovoltaic power converter (PPC) assembly and characterization.} \textbf{a}~Schematic of the fibre-coupled PPC housing. A resin-based 3D-printed housing provides mechanical protection and fixation of the optical fibre, the PPC and their coupling, while a smaller selective-laser-etched (SLE) glass holder ensures precise placement of the TO-can packaged PPC within the housing. The fibre position is adjusted to maximise the PPC output power and then fixed in place with UV-cured glue. \textbf{b}~Current--voltage characteristics of the PPC under 850\,\unit{\nano\meter} illumination at optical powers of 20,40,100, and 200\,\unit{\milli\watt}. The short-circuit current scales with incident optical power while the open-circuit voltage remains near 1.15\,\unit{\volt}, confirming that the converter delivers sufficient current at the LNA supply voltage across the operating range. \textbf{c}~Long-term stability of the PPC output over 1\,\unit{\hour} under constant illumination: the current (top) and voltage (bottom) remain stable to within $\approx$\,2\,\unit{\micro\ampere} and $\approx$\,2\,\unit{\milli\volt}, respectively, demonstrating drift-free optical powering of the on-coil amplifier throughout an imaging session.}
    \label{fig:S3}
\end{figure}

\begin{figure}[h]
    \centering
    \includegraphics[width=0.7\linewidth]{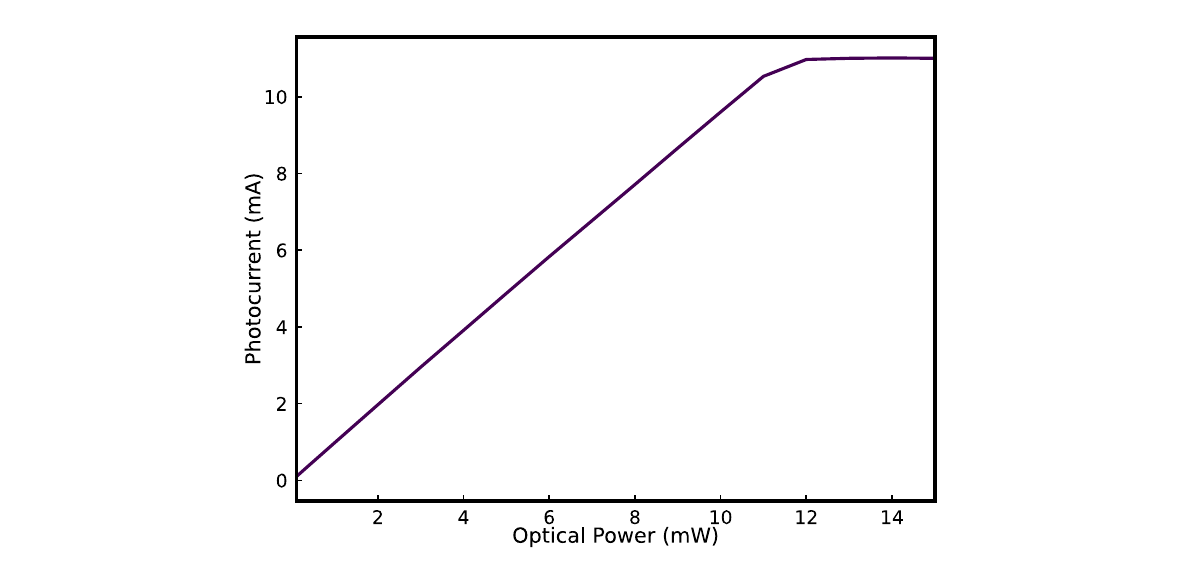}
    \caption{\textbf{Transfer curve of the photodetector.} Photocurrent of the biased InGaAs photodetector versus incident optical power at 1550\,\unit{\nano\meter}. The response is linear up to $\approx$\,11\,\unit{\milli\watt}, where the photocurrent reaches $\approx$\,10.5\,\unit{\milli\ampere}, and saturates at $\approx$\,11\,\unit{\milli\ampere} beyond $\approx$\,12\,\unit{\milli\watt}. The slope of the linear region corresponds to a responsivity of $\mathcal{R} \approx$\,1\,\unit{\ampere\per\watt}. The onset of compression at $\approx$\,11\,\unit{\milli\watt} sets the upper limit of the usable optical operating range of the link.}
    \label{fig:S4}
\end{figure}

\begin{figure}[h]
    \centering
    \includegraphics[width=\linewidth]{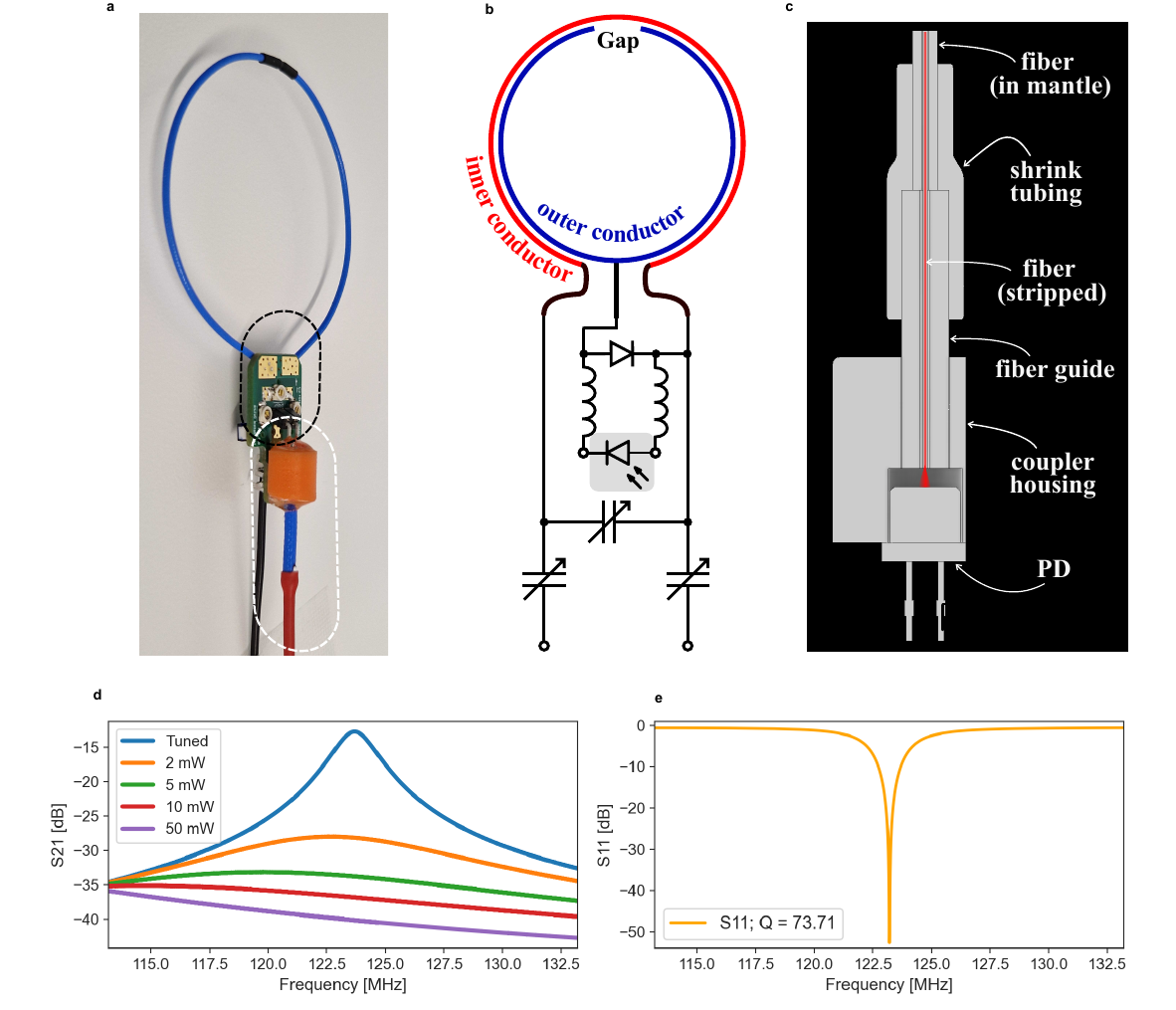}
    \caption{\textbf{Optically controlled detuning of the receive coil.} \textbf{a}~Photograph of the shielded-loop resonator (SLR) used with optical detuning, showing the loop, the matching/detuning board and the fibre-coupled photodiode assembly (white dashed outline). \textbf{b}~Circuit schematic of the coil with the detuning network. During transmission, light delivered to the on-board photodiode (PD; shaded; same device as in \textbf{c}) forward-biases the PIN diode,  which sits in parallel with the tuning capacitor and detunes the loop; the choke inductors separate the AC measurement signal from the DC detuning current. The variable capacitors set tuning to 123.2\,\unit{\mega\hertz} and matching to 50\,\unit{\ohm}. \textbf{c}~Cross-sectional illustration of the mechanical coupling of the fibre to the photodiode: the fibre (in its mantle) is stripped and guided through shrink tubing and a fibre guide into the coupler housing, butt-coupling to the PD (indicated in both \textbf{b} and \textbf{c}). \textbf{d}~Transmission ($S_{21}$) of the coil measured with a double-probe setup for increasing optical power incident on the photodiode. The resonance peak of the tuned coil (blue) is progressively suppressed and shifted as the optical power increases; already 5\,\unit{\milli\watt} shifts the resonance away from the Larmor frequency and effectively detunes the resonator, with near-complete suppression at 50\,\unit{\milli\watt}. \textbf{e}~Reflection ($S_{11}$) of the coil in the tuned state, indicating an unloaded quality factor of $Q = 73.7$.}
    \label{fig:S5}
\end{figure}

\begin{figure}[h]
    \centering
    \includegraphics[width=\linewidth]{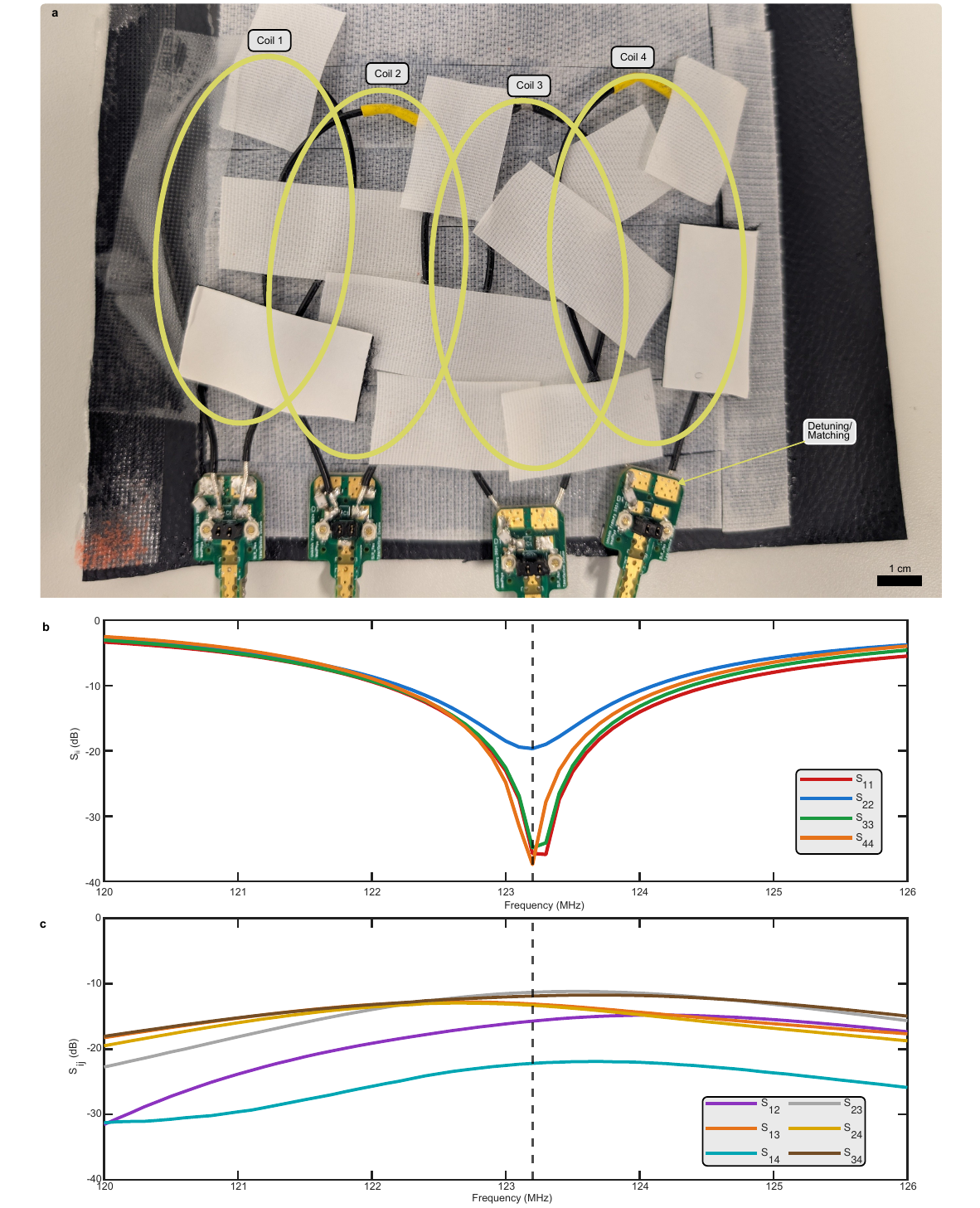}
    \caption{\textbf{Four-channel receive array and its scattering parameters.} \textbf{a}~Photograph of the four-element loop array. The four overlapping coils (Coil~1--4, yellow outlines) are mounted on a leather backing and fixed with hook-and-loop (Velcro) tape, each connected to its own galvanic tuning, matching and detuning board (labelled ``Detuning/Matching''). Scale bar, 1\,\unit{\centi\meter}. \textbf{b}~Reflection coefficients $S_{ii}$ of the four channels, all tuned and matched to the  Larmor frequency at 3\,\unit{\tesla} (123.2\,\unit{\mega\hertz}, dashed line). Each channel reaches a match better than $\approx$-19\,\unit{\decibel} at resonance. \textbf{c}~Coupling coefficients $S_{ij}$ ($i\neq j$) between all coil pairs. Nearest-neighbour coupling is minimised by geometric overlap of adjacent loops, while next-nearest and more distant pairs remain below $\approx$\,-11\,\unit{\decibel} across the band; all inter-element couplings are $\le$\,-11\,\unit{\decibel} at the Larmor frequency.}
    \label{fig:S6}
\end{figure}